\documentclass[aps,pre,10pt,twocolumn]{revtex4-1}
\usepackage{graphicx}
\usepackage{epstopdf}
\usepackage{amsmath}
\usepackage{amssymb}
\usepackage[utf8]{inputenc}
\usepackage{graphicx,float,hyperref}
\usepackage{default}
\usepackage{natbib}
\begin{document}
\title{Optimal Performance of a Three-level Quantum Refrigerator}
%\author{Varinder Singh, Tanmoy Pandit and Ramandeep S. Johal}
%\email{varindersingh@iisermohali.ac.in\\tanmoypandit@iisermohali.ac.in\\rsjohal@iisermohali.ac.in}
\author{Varinder Singh}
\email{varindersingh@iisermohali.ac.in} 
\author{Tanmoy Pandit}
\email{tanmoypandit@iisermohali.ac.in}
\author{Ramandeep S. Johal}
\email{rsjohal@iisermohali.ac.in}
\affiliation{ Department of Physical Sciences, \\ 
Indian Institute of Science Education and Research Mohali,
Sector 81, S.A.S. Nagar, Manauli PO 140306, Punjab, India}
\begin{abstract}
We  study the optimal performance of a three-level quantum refrigerator 
using two different objective 
functions:  cooling power and $\chi$-function. 
For both cases, we obtain general expressions for the 
coefficient of performance (COP) and derive its well-known lower and upper bounds 
for the limiting cases when the ratio
of system-bath coupling constants at the hot and cold contacts
approaches infinity and zero, respectively. 
We also show that the cooling power is optimizable only in the local region
with respect to one control frequency,  
while $\chi$-function can be optimized globally with respect to two control
frequencies. Additionally, we show that in
the low-temperatures regime, our model of refrigerator can be mapped
to Feynman's ratchet and pawl model,
a classical mesoscopic heat engine.
In the parameter regime where both cooling power and $\chi$-function can be
optimized, we compare the cooling power of the quantum refrigerator at maximum 
$\chi$-function with the optimum cooling power.
\end{abstract} 
\maketitle
\section{Introduction}
In 1824, Carnot discovered that the efficiency of any heat
engine operating between two reservoirs at temperatures $T_h$ and $T_c$ ($T_c<T_h$), 
is bounded from above by the Carnot efficiency, 
$\eta_{\rm C}^{}=1-T_c/T_h$. If a heat cycle is 
reversed---turning it into a refrigerator---the corresponding measure,
called the coefficient of performance (COP), is similarly bounded from above 
by $\epsilon_{\rm C}^{} = T_c/(T_h-T_c)$.
Somehow, the optimization analysis of 
 irreversible refrigerators \cite{ YanChen1990, AgarwalMenon, Allahverdyan2010, Apertet2013A}
 turns out to be more involved than that of heat engines.
For instance, power output is a reasonable objective to optimize
for a heat engine.  Under the assumptions of endoreversibility  and Newton's law for heat transfer, 
the efficiency at maximum power was derived by Curzon-Ahlborn (CA) \cite{CA1975}:
\begin{equation}
\eta_{\rm CA} = 1 - \sqrt{1-\eta_C}. \label{ca}
\end{equation}
Then, Esposito and coauthors  \cite{Esposito2010} introduced the concept of a 
low-dissipation heat engine and obtained lower and upper bounds on the 
efficiency  at maximum power. Further, for the symmetric dissipation at the 
hot and the cold contacts, they reproduced CA value. Izumida and
Okuda \cite{IzumidaOkuda} showed that results of low-dissipation model
can be obtained in the optimization of minimally non-linear irreversible heat engines. 
CA-efficiency is also 
obtained using inference in  models of limited information based on Jeffreys 
prior probability function \cite{Sir2010, GJ2015}. 
Recently, in a global approach to irreversible entropy generation \cite{Johal2018} which is independent of 
the specific nature of heat cycle, CA-efficiency was 
related to geometric mean value of the heat exchanged with reservoirs. 

On the other hand, it is not possible to optimize directly the cooling power (CP)
of endoreversible and low-dissipation refrigerators 
in general by taking the same assumptions useful for the optimization 
of a corresponding model of heat engine. 
An expression analogous to CA efficiency was first obtained for refrigerators
by Yan and Chen 
\cite{YanChen1990} by maximizing a new criterion, $\chi=\epsilon \dot{Q}_c$, 
which represents a trade-off between the COP ($\epsilon$) and
 CP ($\dot{Q}_c$) of the refrigerator. The  COP at optimal $\chi$ is given by 
\begin{equation}
\epsilon_{\rm CA}^{}=\sqrt{1+\epsilon_{\rm C}^{}}-1,
\end{equation}
which also holds 
for many models of classical 
\cite{deTomas2012,Izumida2013,Velasco1997,GJ2015,Johal2018,Johal2019} and quantum refrigerators 
\cite{Allahverdyan2010,LutzEPL}.

Agrawal and Menon \cite{AgarwalMenon} showed that CP of endoreversible 
refrigerators becomes optimizable if we take into account the time 
spent on adiabatic branches. However, this results in a model-dependent
expression for the COP. Similarly, CP of a classical endoreversible 
refrigerator can be optimized by considering non-Newtonian laws of heat transfer,
employed earlier to optimize the power output in CA model \cite{YanChen1990}.
Again, this results in non-universal formulae for the COP of the refrigerator 
that depend on phenomenological heat conductivities. Recently, carrying the
research in optimization of refrigerators one step forward, Correa et al.
maximized the CP of a quantum endoreversible refrigerator in high-temperature 
regime and obtained model-independent expression for the COP \cite{Alonso2014B}.

In this work, we study the optimal performance of a three-level quantum 
refrigerator \cite{Scovil1959,Scovil1959B}. 
It is regarded that the study 
of three level systems
pioneered by Scovil and Schulz-DuBois (SSD),  
started the field of quantum thermodynamics \cite{KosloffEntropy,Mahler,
SV2016,Xuereb,DeffnerBook,AlickiKosloff,Binder}. In recent years, these systems have also been 
employed to study quantum heat engines
(refrigerators) \cite{Geva1994,Geva1996,LevyKosloff,Dorfman2018,VJ2019,
Jaseem2019,Uzdin2019}
and quantum absorption refrigerators \cite{Linden2010,Levy2012,Alonso2013,
Bijay2017,Segal2018,Holubec2019, Scarani2019,Mitchison2016,Brunner2015}.
Our choice of the model is 
motivated by the observation that it can be optimized  for both CP and 
$\chi$-function and yields model-independent expressions
for lower and upper bounds on the COP in each case. 

The paper is organized as follows. In Sec.II, we discuss
the model of SSD refrigerator. In
Sec. III, we optimize the CP of the refrigerator and obtain the general
expression for the optimal COP, and find
lower and upper bounds on the COP. In Sec. IV, we optimize the $\chi$-function
and  obtain analytic expressions 
for the COP for global as well as local optimization scheme. We conclude in Sec. V.
\section{Model of Three-Level Quantum Refrigerator}
\begin{figure}   
	\begin{center}
		\includegraphics[width=0.5\textwidth]{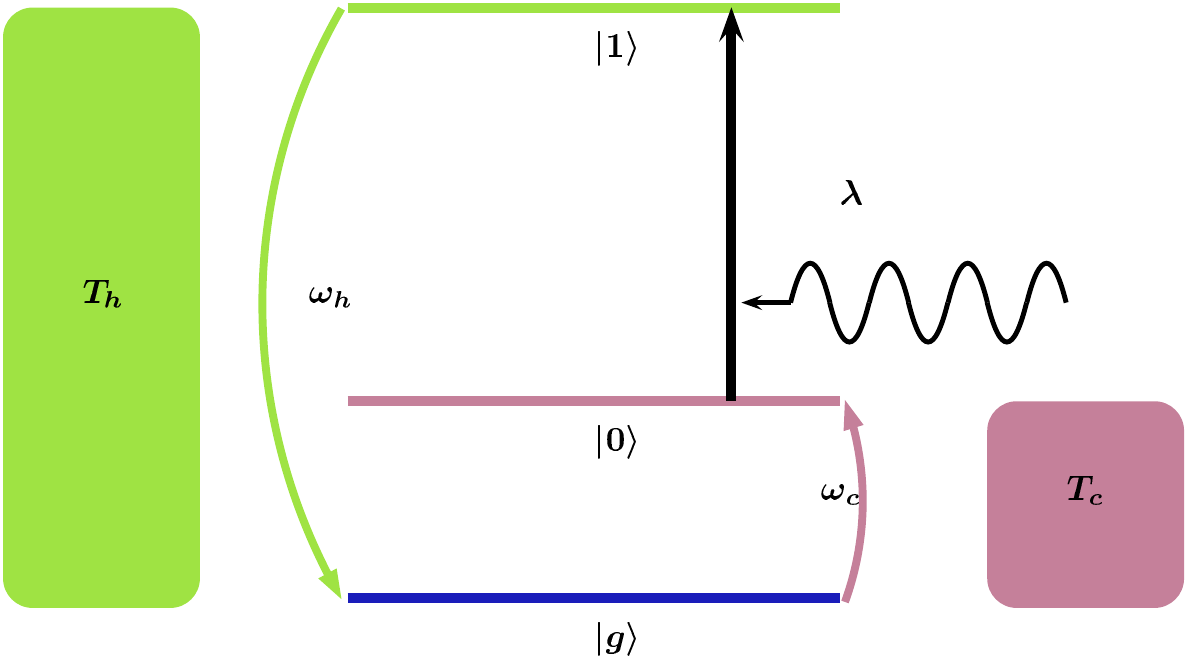}
	\end{center}
	\caption{(Color online) Schematic of three-level laser refrigerator 
	continuously coupled to two heat 
	reservoirs at temperatures $T_c$ and $T_h$ with coupling constants 
	$\Gamma_c$ and $\Gamma_h$
	respectively. A single-mode classical field drives the transition 
	between levels 
	$\vert 0\rangle$ and $\vert 1\rangle$, and $\lambda$ represents the 
	strength of matter-field coupling.} \label{TLL}
\end{figure}

The model consists of a three-level atomic system continuously coupled 
to two thermal reservoirs and
to a single mode of classical electromagnetic field as shown in Fig. \ref{TLL}.
In refrigerators, heat is extracted 
from the cold reservoir and dumped into the hot reservoir, with the help of an 
external agency. The power input
mechanism is modeled by an external single mode field coupled to the levels 
$\vert 0\rangle$ and $\vert 1\rangle$, 
inducing transitions between these levels.
The population in level $\vert 1\rangle$ then relaxes to level $\vert g\rangle$ 
by rejecting heat to the hot bath. The system then jumps from level $\vert g\rangle$
to level $\vert 1\rangle$ by absorbing energy from the cold bath. 
The Hamiltonian of the system is given by: 
$H_0=\hbar \sum \omega_k \vert k\rangle\langle k\vert$, where the summation 
runs over all three states and 
$\omega_k$ represents the relevant atomic frequency. The interaction with the 
single mode lasing field of frequency 
$\omega$, under the rotating wave approximation, is described by the 
semiclassical hamiltonian: 
$V(t)=\hbar \lambda (e^{i\omega t} \vert 1\rangle\langle 0\vert + e^{-i\omega t}
\vert 0\rangle\langle 1\vert)$, where
$\lambda$ is the field-matter coupling constant. The most  general time-independent
dissipator generating a completely positive, trace-preserving and linear evolution 
was derived by Gorini, Kossakowski and Sudarshan \cite{Gorini}, and Lindblad
\cite{Lindblad}. The time  evolution of the system is described by the following 
master equation:
\begin{equation}
\dot{\rho} = -\frac{i}{\hbar} [H_0+V(t),\rho] + \mathcal{L}_{h}[\rho] + 
\mathcal{L}_{c}[\rho],
\end{equation}
where $\mathcal{L}_{h(c)}[\rho]$ represents the dissipative Lindblad superoperator 
describing the system-bath 
interaction with the hot (cold) reservoir:
\begin{eqnarray}
\mathcal{L}_h[\rho] &=& \Gamma_h(n_h+1)(2\vert g\rangle\langle g\vert \rho_{11} - 
\vert 1\rangle\langle 1\vert\rho
- 
\rho\vert 1\rangle\langle 1\vert) \nonumber
\\
&& +\Gamma_h n_h (2\vert 1\rangle\langle 1\vert\rho_{gg}-\vert g\rangle\langle g\vert\rho
-
\rho\vert g\rangle\langle g\vert),\label{dissipator1}
\end{eqnarray} 
\begin{eqnarray}
\mathcal{L}_c[\rho] &=& \Gamma_c(n_c+1)(2\vert g\rangle\langle g\vert \rho_{00} 
- \vert 0\rangle\langle 0\vert\rho
- 
\rho\vert 0\rangle\langle 0\vert) \nonumber
\\
&& +\Gamma_c n_c (2\vert 0\rangle\langle 0\vert\rho_{gg}-\vert g\rangle\langle g\vert\rho
-
\rho\vert g\rangle\langle g\vert).\label{dissipator2}
\end{eqnarray}
Here $\Gamma_h$ and $\Gamma_c$ are the Weisskopf-Wigner decay constants, and 
$n_{h(c)}= 1/(\exp[\hbar\omega_{h(c)}/k_{\rm B} T_{h(c)}]-1)$ is the average 
occupation number of photons 
in hot (cold) reservoir satisfying the relations $\omega_c=\omega_0-\omega_g$,
$\omega_h=\omega_1-\omega_g$.

For our model, it is possible to find a rotating frame in which the steady-state 
density matrix $\rho_R$
is time independent [10]. Defining $\bar{H}=\hbar (\omega_g \vert g\rangle\langle g\vert 
+ \frac{\omega}{2} \vert 1\rangle\langle 1\vert - \frac{\omega}{2} 
\vert 0\rangle\langle 0\vert ) $, 
an arbitrary operator $A$ in the rotating frame is given by 
$A_R=e^{i\bar{H}t/\hbar}Ae^{-i\bar{H}t/\hbar}$. 
It can be seen that $\mathcal{L}_h[\rho]$ and $\mathcal{L}_c[\rho]$ remain 
unchanged under this transformation.
The time evolution of the system density matrix in the rotating frame can be written as
\begin{equation}
\dot{\rho_R} = -\frac{i}{\hbar}[H_0-\bar{H}+V_R,\rho_R] + \mathcal{L}_h[\rho_R] 
+ \mathcal{L}_c[\rho_R]\label{dm1}
\end{equation}
where $V_R=\hbar\lambda(\vert 1\rangle\langle 0\vert + \vert 0\rangle\langle 1\vert)$.

In a series of papers \cite{BoukobzaTannor2006A,BoukobzaTannor2006B,BoukobzaTannor2007},
Boukobza and Tannor formulated a new way of quantifying heat and work for a
weak system-bath coupling \cite{Alicki1979}.
Then, the input power and heat flux of the refrigerator are defined as follows:
\begin{eqnarray}
P &=& \frac{i}{\hbar} {\rm Tr}([H_0,V_R]\rho_R), \label{power1} \\
\dot{Q_c} &=&  {\rm Tr}(\mathcal{L}_c[\rho_R]H_0), \label{heat1} 
\end{eqnarray}
Calculating the traces (see Appendix A) appearing 
in right hand side of the Eqs. (\ref{power1}) and (\ref{heat1}), 
the power and heat flux can be written as:
\begin{equation}
P=i\hbar\lambda(\omega_h-\omega_c)(\rho_{01}-\rho_{10}), \label{power2} 
\end{equation}
\begin{equation}
\dot{Q_c} =  i\hbar\lambda\omega_c (\rho_{10}-\rho_{01}),\label{heat2}
\end{equation}
where $\rho_{01} = \langle 0\vert\rho_R\vert 1\rangle$ and $\rho_{10} =
\langle 1\vert \rho_R\vert 0\rangle$.
Then, the COP is given by
\begin{equation}
\epsilon = \frac{\dot{Q}_c}{P} =\frac{\omega_c}{\omega_h-\omega_c}, \label{efficiency}
\end{equation}
which satisfies $\epsilon \leq \epsilon_{\rm C}^{}$.

%%%%%%%%%%%%%%%%%%%%%%%%%%%%%%%%%%%%%%%%%%%%%%%%%%%%%%%%%%%%%%%%%%%%%%%%%%%%%%%%%%%%%
\section{Optimization Of Cooling Power}
\begin{figure}   
 \begin{center}
\includegraphics[width=8.6cm]{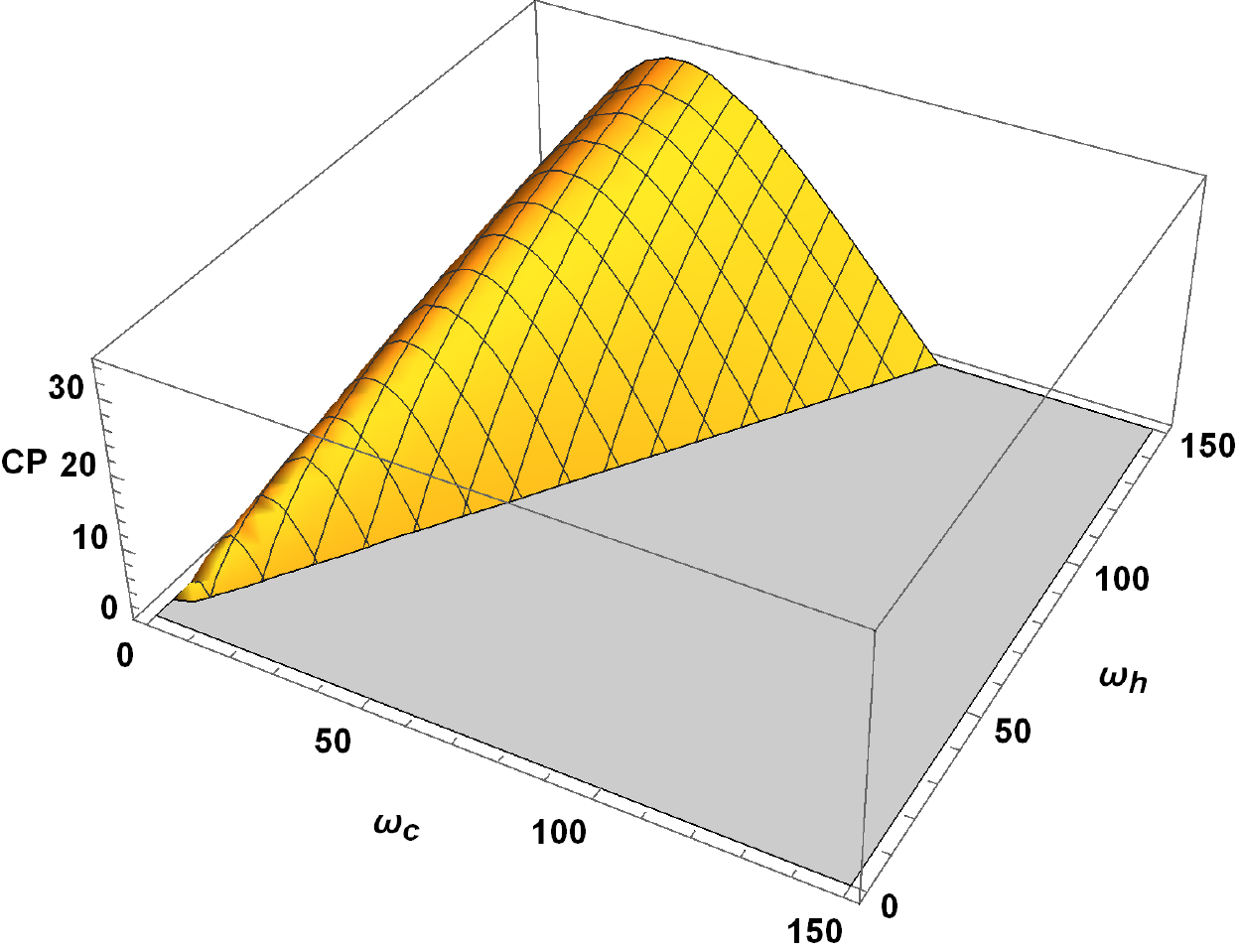}
 \end{center}
\caption{(Color online) 3D-plot of CP [Eq. (\ref{coolpower})] 
in terms of control frequencies $\omega_c$
and $\omega_h$ for $\hbar=1, k_{\rm B}=1, \Gamma_h=3.4, \Gamma_c=3.2, \lambda=3,
T_h=60, T_c=40$.} \label{coolpowerf}
\end{figure}
In this section, we optimize CP of the refrigerator and obtain the expression 
for the corresponding COP.  The general expression for CP  
is derived in Appendix A, see Eq. (\ref{coolpower}). 
We show the 3D-plot of CP with respect to 
$\omega_c$ and $\omega_h$ in Fig. \ref{coolpowerf}. It is clear 
from the figure that a well defined local maximum on
$\omega_c$ exists whereas there is no such local maximum on $\omega_h$.
In other words, CP is optimizable
with respect to $\omega_c$ only. We have played with a wide range of different values
of the concerned parameters ($\Gamma_{c,h}, T_{c,h}, \lambda$), but the basic trend of 
the graph remains the same and it does not change the main result. 
However, in this case, an analytic expression for the COP seems hard to obtain. 

In order to derive the COP in  a closed form, we work in the high-temperatures regime
\cite{Kosloff1984,GevaKosloff1992,Lutz2012,Alonso2013,Alonso2014}
and assume that the matter-field coupling is very strong compared to the system-bath 
coupling ($\lambda \gg \Gamma_{c,h}$) \cite{Dorfman2018}.
In this regime, it is possible to obtain model-independent performance 
benchmarks for both quantum engines and refrigerators 
\cite{Lutz2012,UzdinEPL,Alonso2014B,GevaKosloff1992}. Then,  
we can approximate  
$n_h\simeq k_{\rm B} T_h/\hbar\omega_h$ and $n_c\simeq k_{\rm B} T_c/\hbar\omega_c$ 
and the expression for CP is simplified to be:
\begin{equation}
\dot{Q_c}= 2\hbar\Gamma_h\frac{\omega_c(\tau \omega_h -  \omega_c)}{\tau \omega_h 
+ \gamma \omega_c}, \label{cooll}
\end{equation}
where $\gamma=\Gamma_h/\Gamma_c$ and $\tau=T_c/T_h \equiv \epsilon_{\rm C}^{}/(1+\epsilon_{\rm C}^{})$. 
One can optimize $\dot{Q}_c$ in Eq. (\ref{cooll}) within a local region at fixed
$\omega_h$ by setting $\partial \dot{Q}_c/\partial\omega_c=0$, which
leads to the optimal solution:
\begin{equation}
\omega_c^* = \omega_h \frac{(\sqrt{1+\gamma}-1)\tau}{\gamma}, \label{temp1}
\end{equation}
yielding the following form of the COP at maximum CP 
\begin{equation}
\epsilon^* =\frac{\epsilon_{\rm C}^{}}{1+ \sqrt{1+\gamma}(1+\epsilon_{\rm C}^{})}.
\end{equation}
We note that $\epsilon^*$ is a monotonically decreasing function of $\gamma$. 
Therefore, we can obtain lower and upper bounds on the COP at maximum CP by letting 
$\gamma\rightarrow\infty$ and $\gamma\rightarrow\ 0$, respectively:
\begin{equation}
0 \leq \epsilon^* \leq \frac{\epsilon_{\rm C}^{}}{2+\epsilon_{\rm C}^{}}.
\end{equation}
The above bounds can be obtained in a variety of other models \cite{Izumida2015, Apertet2013A}
and approaches \cite{Johal2018, Johal2019}.
In particular, the upper bound above is also obtained for an endoreversible quantum 
refrigerator (see Eq. (14) in Ref. \cite{Alonso2014B} for $d_c=1$) operating at maximum CP. The reason
behind this is that like  Ref. \cite{Alonso2014B}, we also consider
here the unstructured bosonic baths with a flat spectral density in one-dimension ($d_c=1$).

Similarly, substituting Eq. (\ref{temp1}) in Eq. (\ref{cooll},) the optimal CP is given by:
\begin{equation}
{\dot{Q}_c}^*= 2\hbar\Gamma_h\omega_h\frac{(2+\gamma-2\sqrt{1+\gamma})\epsilon_{\rm C}^{}}
{(1+\epsilon_{\rm C}^{})\gamma^2}.
\end{equation}
For future reference, we find the expressions for ${\dot{Q}_c}^*$ in
the limiting cases $\gamma\rightarrow 0$
and $\gamma\rightarrow \infty$:
\begin{eqnarray}
{\dot{Q}_{c(\gamma\rightarrow 0)}}^* &=& \frac{\hbar\Gamma_h\omega_h}{2}
\frac{\epsilon_{\rm C}^{}}{1+\epsilon_{\rm C}^{}}, \label{E5}
\\
{\dot{Q}_{c(\gamma\rightarrow \infty)}}^* &=& 2\hbar\Gamma_c\omega_h 
\frac{\epsilon_{\rm C}^{}}{1+\epsilon_{\rm C}^{}}. \label{E6}
\end{eqnarray}

\section{ Optimization Of $\chi$-Function}
The $\chi$-function, $\chi=\epsilon\dot{Q}_c$  has already been shown to be 
a suitable figure of merit in the study of optimal performance of classical 
\cite{deTomas2012,Izumida2013} as well as quantum refrigerators 
\cite{Allahverdyan2010,Yuan2014,LutzEPL}.
In the following, we reaffirm this observation by pointing out that in the case 
of  SSD refrigerator, it is possible to 
globally optimize the $\chi$-function with respect to control frequencies 
$\omega_c$ and $\omega_h$. This presents 
the advantage of optimizing $\chi$-function over CP which can only be optimized 
in a local region.
\subsection{Global Optimization}
In the general case, again it is not possible to obtain  analytic expression for the COP. Therefore, 
we optimize Eq. (A12) numerically and present our results in Table I. 
\begin{table}[H]
\caption{COP at global optimization of $\chi$-function.
Here $T_c=50, T_h=100$. The results shown in first, second and third rows correspond to 
$\Gamma_h=1, \Gamma_c=2000; \Gamma_h=1, \Gamma_c=1$; and $\Gamma_h=2000, \Gamma_c=1$, respectively.
For the given values of $T_c$ and $T_h$, $\epsilon_{\rm CA}^{}$ = 0.414213.}
\renewcommand{\arraystretch}{1}
\centering
\begin{tabular}{|c|c|c|c|c|}
\hline
 & $\lambda=1$ &  $\lambda=100$&$\lambda=10000$\\ \hline \hline
$\gamma=0.0005$ & $\epsilon=0.459333$ & $\epsilon=0.475244$&$\epsilon=0.476904$\\ \hline
$\gamma=1$ & $\epsilon=0.441015$ & $\epsilon=0.43729$&$\epsilon=0.437283$\\ \hline    
$\gamma=2000$ & $\epsilon=0.42461$ & $\epsilon=0.372163$&$\epsilon=0.346034$\\ \hline 
\end{tabular}\label{tab:eng-math}
\end{table}
\subsubsection*{Low-temperatures regime}
The low-temperatures regime is governed by the condition:
$k_{\rm B} T_{c,h}\ll\hbar\omega_{c,h}$, such that $n_{c,h}\approx 
e^{-\hbar\omega_{c,h}/k_{\rm B} T_{c,h}}\ll1$.
Simplifying Eq. (A12), we get the expression for $\chi$-function as follows
\begin{equation}
\chi =    \frac{2\hbar\lambda^2 \Gamma_c\Gamma_h(n_c-n_h) \omega_c^2 }
{(\Gamma_c+\Gamma_h)(\lambda^2+\Gamma_c\Gamma_h)(\omega_h-\omega_c)}.  \label{chilow}
\end{equation}
Optimization of Eq. (\ref{chilow}), with respect
to $\omega_h$ and $\omega_c$, yields the following equations:
\begin{eqnarray}
e^{\hbar\omega_h/k_{\rm B} T_h-\hbar\omega_c/k_{\rm B} T_c} &=& 1 
+ \frac{\hbar\omega_c\epsilon_{\rm C}^{}}{k_{\rm B} T_c\epsilon(1+\epsilon_{\rm C}^{})}, 
\\
e^{\hbar\omega_h/k_{\rm B} T_h-\hbar\omega_c/k_{\rm B} T_c} &=& \frac{k_{\rm B} T_c(2+\epsilon)}
{k_{\rm B} T_c(2+\epsilon)-\hbar\omega_c}.
\end{eqnarray}
The above equations cannot be solved analytically for $\omega_h$ and $\omega_c$.
However, they can be combined to 
give the following transcendental equation:
\begin{equation}
\frac{(\epsilon_{\rm C}^{}-\epsilon)(2\epsilon_{\rm C}^{}-\epsilon)}{\epsilon\epsilon_{\rm C}^{}(1+\epsilon_{\rm C}^{})}
 =
 \ln\left[\frac{(2+\epsilon)\epsilon_{\rm C}^{}}{\epsilon(1+\epsilon_{\rm C}^{})}\right], \label{trans}
\end{equation}
which clearly indicates that COP at maximum $\chi$-function depends 
upon $\epsilon_{\rm C}^{}$ only and is independent
of system parameters. Eq. (\ref{trans}) along with the expression, 
$\epsilon_{\rm CA}^{}=\sqrt{1+\epsilon_{\rm C}^{}}-1$,  
is plotted in Fig. 3, from which it is clear that COP of the SSD refrigerator
operating in low-temperatures regime is higher than, though quite close to $\epsilon_{\rm CA}^{}$. 
See also Appendix D for the mapping of the refrigerator model in the above regime
to Feynman's ratchet and pawl model. 
\begin{figure}   
 \begin{center}
\includegraphics[width=8.6cm]{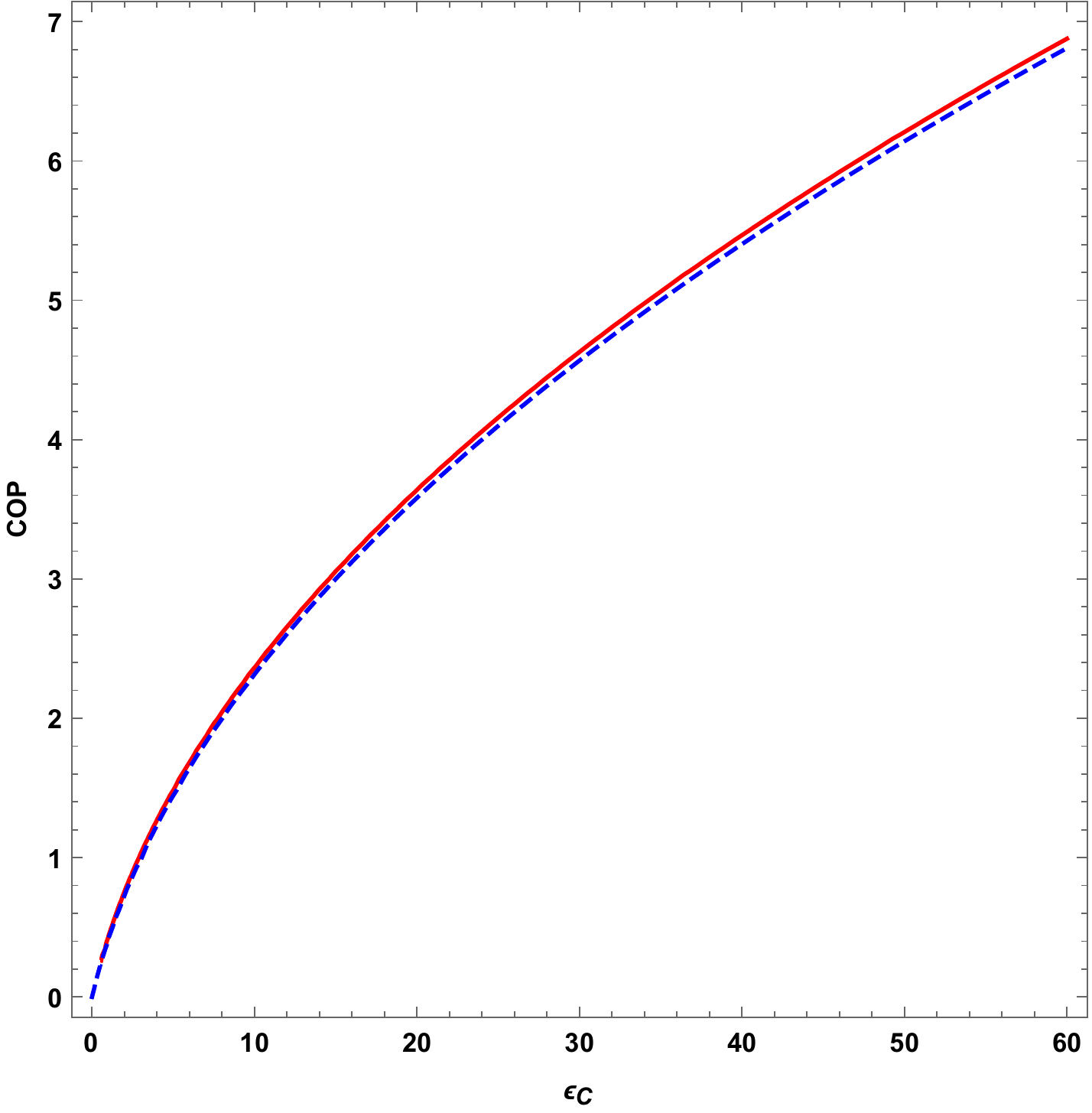}
 \end{center}
\caption{(Color online) Plot of the COP versus $\epsilon_{\rm C}^{}$. 
Solid red curve represents Eq. (\ref{trans}) and
dashed blue curve represents the equation $\epsilon_{\rm CA}^{}=\sqrt{1+\epsilon_{\rm C}^{}}-1$.}  
\end{figure}
\subsection{Local optimization in high-temperatures regime}
High temperatures along with a strong matter-field coupling is another operational
regime in which we can obtain model-independent benchmarks from the optimization of $\chi$-function. 
In this regime, the expression for $\chi$ is simplified to:
\begin{equation}
\chi = \epsilon \dot{Q_c}= \frac{2\hbar\Gamma_h\omega_c^2(\tau \omega_h -  \omega_c)}
{(\tau \omega_h + \gamma \omega_c)
(\omega_h-\omega_c)}. \label{chicool}
\end{equation}
If we attempt a two-parameter optimization by setting $\partial\chi/\partial\omega_c=0$
and $\partial\chi/\partial\omega_h=0$, it gives the trivial solution, $\omega_h=\omega_c=0$.
Although in the previous section, 
we have shown the existence of global maximum of $\chi$ under general conditions, 
no such global maximum exists in this regime. It indicates that the 
assumption of high temperatures might not be justified for the simultaneous 
optimization with respect to $\omega_c$ and $\omega_h$.
Since two-parameter optimization fails, we optimize $\chi$-function alternately with 
respect to $\omega_h$ ($\omega_c$ fixed)  and 
$\omega_c$ ($\omega_h$ fixed). For fixed $\omega_c$, setting 
$\partial\chi/\partial\omega_h=0$, we obtain 
\begin{equation}
\omega_h = \omega_c \frac{\gamma-\tau(1+\gamma)}{ \tau \big(1-\sqrt{(1+\gamma)(1-\tau)}\big)}.
\label{aux1}
\end{equation}
Substituting in Eq. (\ref{efficiency}), and writing in terms of Carnot  COP $\epsilon_{\rm C}^{}$, 
we get the following form of COP at maximum $\chi$-function:
\begin{equation}
\epsilon^* = \frac{\epsilon_{\rm C}^{}}{1+\sqrt{(1+\gamma)(1+\epsilon_{\rm C}^{})}}.
\end{equation}
Again $\epsilon^*$ is monotonic decreasing function of $\gamma$. Therefore, 
we can obtain lower and upper bounds on
the COP by putting $\gamma\rightarrow\infty$ and $\gamma\rightarrow 0$, respectively:
\begin{equation}
\epsilon_- \equiv 0 \leq \epsilon^* \leq  \epsilon_{\rm CA}^{}.
\end{equation}
The lower bound, $\epsilon_-=0$, concurs with the lower bound of low-dissipation 
\cite{WangLi2012} 
and minimally non-linear irreversible models of refrigerators \cite{Izumida2013}.
As mentioned earlier, the upper bound, $\epsilon_{\rm CA}^{}$, 
was first derived for a classical endoreversible refrigerator \cite{YanChen1990}. 
Under the conditions of tight-coupling and
symmetric dissipation, $\epsilon_{\rm CA}^{}$ can also be obtained for the
low-dissipation \cite{deTomas2012} and minimally non-linear irreversible refrigerators 
\cite{Izumida2013}. For a quantum Otto refrigerator, the COP emerges  
to be equal to $\epsilon_{\rm CA}^{}$ in the classical limit (high-temperatures limit)
\cite{LutzEPL}. 

%%%%%%%%%%%%%%%%%%%%%%%%%%%%%%%%%%%%%%%%%%%%%%%%%%%%%%%%%%%%%%%%%%%%%%%%%%%%%%%%%%%%%%%%%%%%%5
Next, we optimize $\chi$ with respect to $\omega_c$ while keeping $\omega_h$ constant. 
In this case, $\partial\chi / \partial\omega_c$=0, yields the following equation:
\begin{widetext}
\begin{equation}
\frac{\omega_c[{\gamma\omega_c^3 + 2\omega_h(\tau-\gamma)\omega_c^2
- \tau\omega_h^2(3+\tau - \gamma)\omega_c 
+
2\tau^2\omega_h^3}]}{(\omega_c-\omega_h)^2(\gamma\omega_c+\tau\omega_h)} =0.
\label{aux2}
\end{equation}
\end{widetext}
Due to Casus irreducibilis (see Appendix C), the roots of the cubic equation 
inside the square brackets above 
can only be expressed using complex 
radicals, although the roots are actually real. We can still obtain the 
lower and upper bounds on the COP by solving 
Eq. (\ref{aux2}) for the limiting cases $\gamma\rightarrow \infty$ and 
$\gamma\rightarrow 0$, respectively. An
alternative method is explained in Appendix B that obtains the same 
expressions.
For $\gamma\rightarrow\infty$, the COP is evaluated at CA value. For
$\gamma\rightarrow 0$, we obtain the upper bound on the COP as 
$\epsilon_+=(\sqrt{9+8\epsilon_{\rm C}^{}}-3)/2$. 
Further, although we cannot see analytically, numerical evidence
shows that COP lies
in the range:
\begin{equation}
\epsilon_{\rm CA}^{} \leq \epsilon^* \leq \frac{1}{2} (\sqrt{9+8\epsilon_{\rm C}^{}}-3)
\equiv \epsilon_+.
\end{equation}
Interestingly, $\epsilon_{\rm CA}^{}$ also appears as the 
lower bound for the optimization of a quantum model of refrigerator 
consisting of two $n$-level systems interacting
via a pulsed external field \cite{Allahverdyan2010}. However, the result 
reported in Ref. \cite{Allahverdyan2010} was obtained in the linear response
regime where $T_c\approx T_h$. In the same model, imposing the condition of 
equidistant spectra, $\epsilon_{\rm CA}^{}$ can 
be obtained as an upper bound in the classical regime  for $n\rightarrow\infty$.
The upper bound
$\epsilon_+=(\sqrt{9+8\epsilon_{\rm C}^{}}-3)/2$ obtained here also serves as the upper 
limit on the COP for low-dissipation
\cite{WangLi2012} and minimally non-linear irreversible models \cite{Izumida2013}.
Further, for a two-level quantum system working as a refrigerator, the same upper
bound can be derived in the high temperature regime \cite{Yuan2014}.  
\begin{figure}   
 \begin{center}
\includegraphics[width=8.6cm]{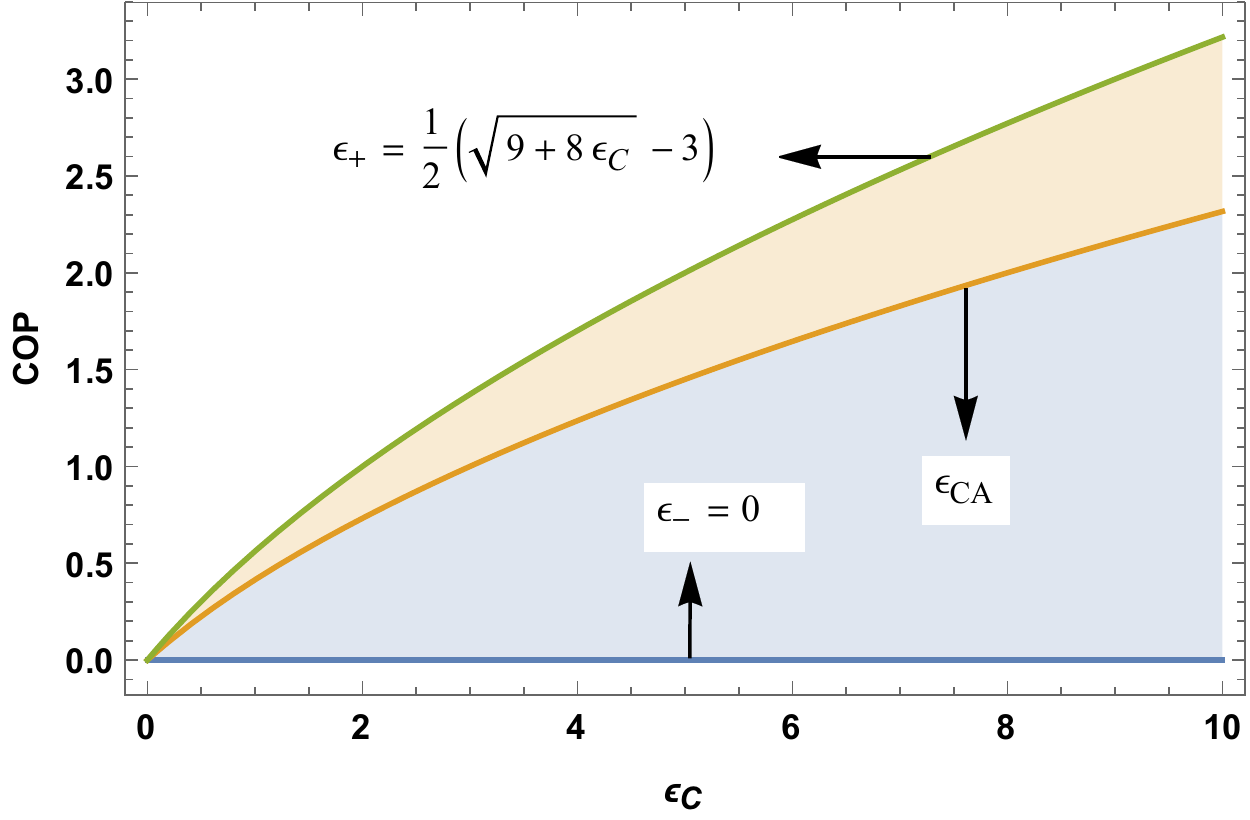}
 \end{center}
\caption{(Color online) Plot of the COP at optimal $\chi$ versus $\epsilon_{\rm C}^{}$.
$\epsilon_{\rm CA}^{}$ divides the parametric region of the COP into two parts. 
For the optimization of $\chi$-function 
over $\omega_h$, it serves as an upper bound, whereas it is the lower bound 
on COP for the optimization over $\omega_c$.}  
\end{figure}

\section{Cooling power at optimal $\chi$-function versus 
optimal cooling power}
In this section, we  compare the CP obtained at maximum $\chi$-function 
with the optimal CP.
As CP can be optimized with respect to $\omega_c$ only, 
we can make the comparison only for this case. 
Dividing Eq. (\ref{E55}) by Eq. (\ref{E5}), we get the ratio of CP at
maximum $\chi$-function
to the optimal CP, for the limiting case $\gamma\rightarrow 0$:
\begin{equation}
R_{\gamma\rightarrow 0} = \frac{(3+2\epsilon_{\rm C}^{})\sqrt{9+8\epsilon_{\rm C}^{}} 
- 9 -10\epsilon_{\rm C}^{}}{2\epsilon_{\rm C}^{2}}, \label{R1}
\end{equation}
which approaches the value 8/9 for small $\epsilon_{\rm C}^{}$, while it vanishes for 
large $\epsilon_{\rm C}^{}$ (small temperature differences).

\begin{figure}   
 \begin{center}
\includegraphics[width=8.6cm]{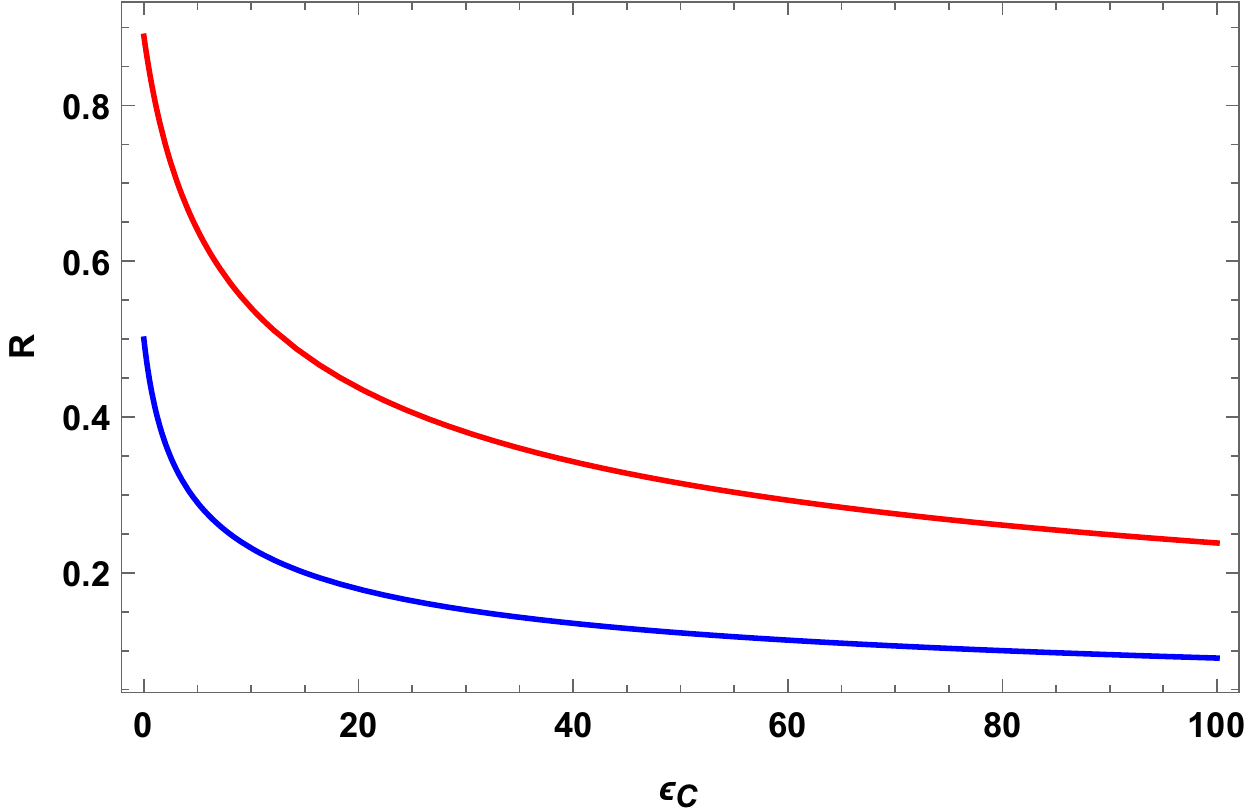}
 \end{center}
\caption{(Color online) Ratio ($R$) of the CP at optimal $\chi$-function to the optimal CP.
Red and blue curves represent Eqs. (\ref{R1}) and (\ref{R2}), respectively,
which approach the value 8/9 and 1/2 respectively for $\epsilon_{\rm C}^{}\rightarrow 0$,
and vanish for $\epsilon_{\rm C}^{}\rightarrow\infty$.}  
\label{figratio}
\end{figure}

Similarly, we get the corresponding ratio for $\gamma\rightarrow\infty$ upon 
dividing Eq. (\ref{E66}) by Eq. (\ref{E6}):
\begin{equation}
R_{\gamma\rightarrow\infty} = \frac{\sqrt{1+\epsilon_{\rm C}^{}}-1}{\epsilon_{\rm C}^{}}, \label{R2}
\end{equation}
which approaches the value 1/2 for small $\epsilon_{\rm C}^{}$, while it vanishes for 
large $\epsilon_{\rm C}^{}$.
We have plotted Eqs. (\ref{R1}) and (\ref{R2}) in Fig. \ref{figratio}, 
from which it is clear that the ratio is greater
for the case $\gamma\rightarrow 0$. Further, it is interesting to note 
that although both $R_{\gamma\rightarrow 0}$ and 
$R_{\gamma\rightarrow\infty}$ vanish for $\epsilon_{\rm C}^{}\rightarrow\infty$,
their ratio 
${R_{\gamma\rightarrow 0}}/{R_{\gamma\rightarrow\infty}} \to 2\sqrt{2}$ 
for small temperature
differences.

\section{Conclusions}
In this work, we have studied the optimal performance of a three-level atomic
system working as a refrigerator. 
We have studied two different target functions: CP and $\chi$-function.
Although, in many classical and quantum models of the refrigerator, CP is not 
a good figure of merit to optimize,  
in our model, it is a well-behaved function and we have obtained analytic 
expressions for lower and upper bounds on the COP already
derived in some models of classical and quantum refrigerators. However, 
we notice that  CP is optimizable
only with respect to the control frequency $\omega_c$ and thus, we can 
perform  optimization in local region of the parameter space only. 
In contrast to the behavior of CP, $\chi$-function shows global maximum 
which makes it 
a more suitable figure of merit to study the optimal performance of 
refrigerators. In the general unconstrained regime, we have presented 
results of numerical optimization in Table I. 
Then in the low-temperatures regime, we showed that the COP 
of our model is independent of system-bath coupling ($\Gamma_{c,h}$) 
or matter-field coupling ($\lambda$), 
and depends on Carnot COP only, which is a remarkable result.
Further, in the high temperature and strong-coupling  
regime, we have 
alternatively performed maximization of $\chi$-function with respect 
to $\omega_h$ ($\omega_c$ fixed) and
$\omega_c$ ($\omega_h$ fixed). In both cases, we were able to obtain 
the lower and upper limits on the COP,
already well known in the  literature on optimization of refrigerators. 
 The possibility of 
simultaneous optimization of CP and $\chi$-function, enables a comparison
between optimal CP in the quantum refrigerator with the CP at optimal $\chi$-function,
and we conclude that a large system-bath coupling at the cold end (compared
to the hot end) yields a higher relative value of CP (see Fig. 5).
There are a few classical models \cite{Apertet2013A, Izumida2013, Izumida2015}
in which both CP and $\chi$-function are optimizable. To the best 
of our knowledge, the present model provides an instance of a quantum
thermal machine allowing the same feature. This will aid future studies \cite{Benjamin}
which explore models in which the performance of quantum machines
can be bettered over their classical counterparts. 
\section*{Acknowledgments}
The authors gratefully acknowledge useful discussions with Sibasish Ghosh.
\appendix\section{Steady state solution  of density matrix equations}
Here, we solve the equations for density matrix in the steady state. 
Substituting the expressions for $H_0$, $\bar{H}$,
$V_0$, and using Eqs. (\ref{dissipator1}) and (\ref{dissipator2}) in Eq. (\ref{dm1}), the time evolution
of the elements of the density matrix are given by following equations:
\begin{eqnarray}
\dot{\rho}_{11} &=& i\lambda (\rho_{10}-\rho_{01}) - 2\Gamma_h[(n_h+1)
\rho_{11}-n_h\rho_{gg}],\label{A1} \\
\dot{\rho}_{00} &=& -i\lambda (\rho_{10}-\rho_{01}) - 2\Gamma_c[(n_c+1)
\rho_{00}-n_c\rho_{gg}], \\
\dot{\rho}_{10} &=& -[\Gamma_h(n_h+1)+\Gamma_c(n_c+1)]\rho_{10} 
+ i\lambda(\rho_{11}-\rho_{00}), \nonumber
\\
\\
\rho_{11} &= & 1-\rho_{00} - \rho_{gg}, \\
\dot{\rho}_{01} &=& \dot{\rho}_{10}^*. \label{A5}
\end{eqnarray}
Solving Eqs. (\ref{A1}) - (\ref{A5}) in the steady state by setting 
$\dot{\rho}_{mn}=0$ ($m,n=0,1$), we obtain
\begin{widetext}
	\begin{equation}
	\rho_{10} = \frac{i\lambda(n_h-n_c)\Gamma_c\Gamma_h}
	{
		\lambda^2[(1+3n_h)\Gamma_h + (1+3n_c)\Gamma_c] + \Gamma_c
		\Gamma_h[1+2n_h+n_c(2+3n_h)][(1+n_c)\Gamma_c + (1+n_h)\Gamma_h ] 
	},\label{rho10}
	\end{equation}
\end{widetext}
and 
\begin{equation}
\rho_{01} = \rho_{10}^*. \label{rho01}
\end{equation}
Calculating the trace in Eq. (\ref{power1}), the input power is given by
\begin{equation}
P = i\hbar\lambda(\omega_h-\omega_c)(\rho_{10}-\rho_{01}). \label{power3}
\end{equation}
Similarly evaluating the trace in Eq. (\ref{heat1}), heat flux $\dot{Q}_c$ can be written as
\begin{equation}
\dot{Q}_c = \hbar \omega_c (2\Gamma_c [n_c\rho_{gg}-(n_c+1)\rho_{00}]). \label{heat3}
\end{equation}
Using the steady state condition $\dot{\rho}_{00}=0$ (see Eq. (\ref{A1})), Eq. (\ref{heat3}) becomes
\begin{equation}
\dot{Q}_c = i\hbar \lambda\omega_c(\rho_{10}-\rho_{01}). \label{aux4}
\end{equation}

Substituting Eqs. (\ref{rho10}) and (\ref{rho01}) in  Eq. (\ref{aux4}), we have
\begin{widetext}
	\begin{equation}
	\dot{Q_c} =    \frac{2\hbar\lambda^2 \Gamma_c\Gamma_h(n_c-n_h) \omega_c }
	{
		\lambda^2[(1+3n_h)\Gamma_h + (1+3n_c)\Gamma_c] + \Gamma_c\Gamma_h
		[1+2n_h+n_c(2+3n_h)][(1+n_c)\Gamma_c + (1+n_h)\Gamma_h] 
	}. \label{coolpower}
	\end{equation}
	 The expression for $\chi$-function, $\chi=\epsilon  \dot{Q_c}$, is given by
	\begin{equation}
	\chi =    \frac{2\hbar\lambda^2 \Gamma_c\Gamma_h(n_c-n_h) \omega_c^2 }
	{
		\lambda^2(\omega_h-\omega_c)[(1+3n_h)\Gamma_h + (1+3n_c)\Gamma_c] 
		+ \Gamma_c\Gamma_h[1+2n_h+n_c(2+3n_h)][(1+n_c)\Gamma_c + (1+n_h)\Gamma_h] 
	}. \label{chipower}
	\end{equation}
\end{widetext}
For refrigerator, $n_c>n_h$, and thus $\dot{Q}_c, \chi > 0$.

\section{Optimization of $\chi$-function with respect to $\omega_c$
in high-temperature and strong coupling regime}
The expression for the $\chi$-function is given by
\begin{equation}
\chi = \frac{2\hbar\Gamma_h\omega_c^2(\tau\omega_h-\omega_c)}{(\tau\omega_h
+\gamma\omega_c)(\omega_h-\omega_c)}.
\end{equation}
As explained in the Section IV, we cannot optimize the above function with 
respect to $\omega_c$ to obtain the roots in real radicals because of the 
Casus irreducibilis (see Appendix C). However, we can obtain the real 
solutions for the limiting 
cases $\gamma\rightarrow 0$ and $\gamma\rightarrow\infty$. 
For $\gamma\rightarrow 0$, $\chi$-function can be written as
\begin{equation}
\chi = \frac{2\hbar\Gamma_h\omega_c^2(\tau\omega_h-\omega_c)}{\tau
\omega_h(\omega_h-\omega_c)},\label{E1}
\end{equation}
which can be optimized to give
\begin{equation}
\omega_c = \frac{\omega_h}{4} \Big(3+\tau-\sqrt{9-10\tau+\tau^2}\Big). \label{E2}
\end{equation}
Substituting Eq. (\ref{E2}) in Eq. (\ref{E1}) and in the equation 
$\dot{Q_c}=2\hbar\Gamma_h\omega_c(\tau\omega_h-\omega_c)/\tau\omega_h$, 
we get following expressions for the optimal
$\chi$-function and CP at optimal $\chi$-function, respectively:
\begin{eqnarray}
\chi^{*(\omega_h)}_{\gamma\rightarrow 0} &=& 
\hbar\Gamma_h\omega_h \frac{27+36\epsilon_{\rm C}^{}+8\epsilon_{\rm C}^{2}-(9+8\epsilon_{\rm C}^{})^{3/2}}
{4\epsilon_{\rm C}^{}(1+\epsilon_{\rm C}^{})},  \nonumber
  \\ \label{E3} \\
\dot{Q_c}^{\chi(\omega_h)}_{\gamma\rightarrow 0} &=&  
\hbar\Gamma_h\omega_h\frac{(3+2\epsilon_{\rm C}^{})\sqrt{9+8\epsilon_{\rm C}^{}}- 9 -10\epsilon_{\rm C}^{}}
{4\epsilon_{\rm C}^{}(1+\epsilon_{\rm C}^{})}. \nonumber
\\ \label{E55}
\end{eqnarray}
Similarly, for $\gamma\rightarrow\infty$, optimization of $\chi$-function, 
$\chi=2\hbar\Gamma_c(\tau\omega_h-\omega_c)/(\omega_h-\omega_c)$, yields 
the following expressions:
\begin{eqnarray}
\chi^{*(\omega_h)}_{\gamma\rightarrow\infty} &=& 
\frac{2\hbar\Gamma_c\omega_h(2+\epsilon_{\rm C}^{}-\sqrt{1+\epsilon_{\rm C}^{}})}{1+\epsilon_{\rm C}^{}},
\label{E4}
\\\nonumber
\\ 
\dot{Q_c}^{\chi(\omega_h)}_{\gamma\rightarrow\infty} &=& \frac{2\hbar\Gamma_c
\omega_h(\sqrt{1+\epsilon_{\rm C}^{}}-1)}
{1+\epsilon_{\rm C}^{}}.\label{E66}
\end{eqnarray}
\section{Casus Irreducibilis}
In algebra, Casus irreducibilis arises while solving a cubic equation.
The formal statement of the Casus irreducibilis 
is that if a cubic polynomial is irreducible with rational coefficients 
and has three real roots, then the roots of 
the cubic equation are not expressible using real radicals and thus, 
one must introduce expressions with complex radicals,
even though the resulting expressions are actually real-valued.
It was proven by P. Wantzel in 1843 \cite{Kleiner2007}.  Using the 
discriminant $D$ of the irreducible cubic equation, 
one can decide whether the given equation is in Casus irreducibilies or not,
via Cardano's formula \cite{Stewart1990}.
The most general form of a cubic equation is given by 
\begin{equation}
a x^3 + b x^2 + c x + d  =0
\end{equation}
where $a, b, c, d$ are real.

The discriminant $D$ is given by: $D=18abcd-4b^3d+b^2c^2-4ac^3-27a^2d^2$. 
Depending upon the sigh of
$D$, following three cases arise:
\\
(a) $ D < 0$, the cubic equation has two complex  roots, so 
Casus irresucibilies does not apply.
\\
(b) $D=0$, all three roots are real and expressible by real radicals. 
\\
(c) $D > 0$, three are three distinct  real roots. In this case, a rational 
root exists and can be found using 
the rational root test. Otherwise, the given polynomial is Casus irreducibilis 
and we need complex valued expressions to 
express the roots in radicals.

In our case, in order to solve Eq. (\ref{aux2}), we have to solve the following cubic equation
\begin{widetext}
\begin{equation}
\gamma\omega_c^3 + 2\omega_h(\tau-\gamma)\omega_c^2- \tau\omega_h^2(3+\tau - \gamma)\omega_c 
+
2\tau^2\omega_h^3 =0. \label{Casus}
\end{equation}
The discriminant $D$ of the above equation is given by
\begin{equation}
D = 4\omega_h^6 (1 + \gamma) (1 + \tau)
[3 \gamma^2 (3 -  \tau) + \gamma^3 + 9 \gamma \tau +  3 \gamma\tau^2 + 9 \tau^2 (1 - \tau)].
\end{equation}
\end{widetext}
Since the parameters $\omega_h, \gamma, \tau$ are positive and $\tau<1$, $D>0$. 
So the polynomial in Eq. (\ref{Casus})
presents the case of Casus irreducibilis.

\section{Mapping to Feynman's ratchet and pawl model}
It is interesting to note that in the low-temperatures regime, SSD refrigerator can 
be mapped to Feynman's ratchet and pawl model 
\cite{Feynman,Parrondo,ShengYangFR,VarinderJohal,VarinderJohal2018}, 
a mesoscopic steady-state heat engine capable of extracting work from thermal 
fluctuations from a
setup of two heat reservoirs via a ratchet and pawl  mechanism. 
In the refrigerator mode, the ratchet makes a backward jump
when $x_c$ amount of heat is absorbed from the cold reservoir and subsequently  
$x_h$ amount of heat
is supplied to the hot reservoir \cite{Feynman,VarinderJohal}. Similarly, the wheel 
turns in the forward direction when 
$x_h$ energy is absorbed from the hot reservoir. The rates of forward and backward 
jumps are given by
\begin{equation}
R_F = r_0 e^{-\hbar x_h/k_{\rm B} T_h},\qquad R_B = r_0 e^{-\hbar x_c/k_{\rm B} T_c}, 
\end{equation} 
where $r_0$ is the rate constant. The system operates as a refrigerator when 
$R_B>R_F$. The rates of heat exchanged with the cold and hot reservoirs, 
respectively, are given by
\begin{eqnarray}
\dot{Q}_c &=& x_c(R_B-R_F) = r_0 x_c (e^{-\hbar x_c/k_{\rm B} T_c} 
- e^{-\hbar x_h/k_{\rm B} T_h}),\nonumber
\\
\\
\dot{Q}_h &=& x_h(R_B-R_F) = r_0 x_h (e^{-\hbar x_c/k_{\rm B} T_c} 
- e^{-\hbar x_h/k_{\rm B} T_h}).\nonumber
\\
\end{eqnarray}
Therefore, $\chi$-function for Feynman's model can be written as follows
\begin{equation}
\chi_{\rm F}^{} = \frac{r_0 x_c^2 }{x_h- x_c}
\left(e^{-\hbar x_c/k_{\rm B} T_c} - e^{-\hbar x_h/k_{\rm B} T_h}\right). \label{chiFR}
\end{equation}
Apart from the multiplicative constant 
$2\hbar\lambda^2\Gamma_c\Gamma_h/(\Gamma_c+\Gamma_h)(\lambda^2+\Gamma_c\Gamma_h)$
(instead of $r_0$), 
the expression in Eq. (\ref{chilow}) is similar to the $\chi$-function 
for the Feynman's model [Eq.(\ref{chiFR})], where 
$\omega_c$ and $\omega_h$ are replaced by $x_c$ and $x_h$. Thus, 
we establish a mapping between our 
model of refrigerator and Feynman's model.  A similar mapping also exists
between the SSD engine and Feynman's ratchet as heat engine \cite{VJ2019}.

%\bibliography{biblo}

\begin{thebibliography}{63}%
\makeatletter
\providecommand \@ifxundefined [1]{%
 \@ifx{#1\undefined}
}%
\providecommand \@ifnum [1]{%
 \ifnum #1\expandafter \@firstoftwo
 \else \expandafter \@secondoftwo
 \fi
}%
\providecommand \@ifx [1]{%
 \ifx #1\expandafter \@firstoftwo
 \else \expandafter \@secondoftwo
 \fi
}%
\providecommand \natexlab [1]{#1}%
\providecommand \enquote  [1]{``#1''}%
\providecommand \bibnamefont  [1]{#1}%
\providecommand \bibfnamefont [1]{#1}%
\providecommand \citenamefont [1]{#1}%
\providecommand \href@noop [0]{\@secondoftwo}%
\providecommand \href [0]{\begingroup \@sanitize@url \@href}%
\providecommand \@href[1]{\@@startlink{#1}\@@href}%
\providecommand \@@href[1]{\endgroup#1\@@endlink}%
\providecommand \@sanitize@url [0]{\catcode `\\12\catcode `\$12\catcode
  `\&12\catcode `\#12\catcode `\^12\catcode `\_12\catcode `\%12\relax}%
\providecommand \@@startlink[1]{}%
\providecommand \@@endlink[0]{}%
\providecommand \url  [0]{\begingroup\@sanitize@url \@url }%
\providecommand \@url [1]{\endgroup\@href {#1}{\urlprefix }}%
\providecommand \urlprefix  [0]{URL }%
\providecommand \Eprint [0]{\href }%
\providecommand \doibase [0]{http://dx.doi.org/}%
\providecommand \selectlanguage [0]{\@gobble}%
\providecommand \bibinfo  [0]{\@secondoftwo}%
\providecommand \bibfield  [0]{\@secondoftwo}%
\providecommand \translation [1]{[#1]}%
\providecommand \BibitemOpen [0]{}%
\providecommand \bibitemStop [0]{}%
\providecommand \bibitemNoStop [0]{.\EOS\space}%
\providecommand \EOS [0]{\spacefactor3000\relax}%
\providecommand \BibitemShut  [1]{\csname bibitem#1\endcsname}%
\let\auto@bib@innerbib\@empty
%</preamble>
\bibitem [{\citenamefont {Yan}\ and\ \citenamefont {Chen}(1990)}]{YanChen1990}%
  \BibitemOpen
  \bibfield  {author} {\bibinfo {author} {\bibfnamefont {Z.}~\bibnamefont
  {Yan}}\ and\ \bibinfo {author} {\bibfnamefont {J.}~\bibnamefont {Chen}},\
  }\href@noop {} {\bibfield  {journal} {\bibinfo  {journal} {J. Phys. D: Appl.
  Phys.}\ }\textbf {\bibinfo {volume} {23}},\ \bibinfo {pages} {136} (\bibinfo
  {year} {1990})}\BibitemShut {NoStop}%
\bibitem [{\citenamefont {Agrawal}\ and\ \citenamefont
  {Menon}(1990)}]{AgarwalMenon}%
  \BibitemOpen
  \bibfield  {author} {\bibinfo {author} {\bibfnamefont {D.~C.}\ \bibnamefont
  {Agrawal}}\ and\ \bibinfo {author} {\bibfnamefont {V.~J.}\ \bibnamefont
  {Menon}},\ }\href@noop {} {\bibfield  {journal} {\bibinfo  {journal} {J.
  Phys. A}\ }\textbf {\bibinfo {volume} {23}},\ \bibinfo {pages} {5319}
  (\bibinfo {year} {1990})}\BibitemShut {NoStop}%
\bibitem [{\citenamefont {Allahverdyan}\ \emph {et~al.}(2010)\citenamefont
  {Allahverdyan}, \citenamefont {Hovhannisyan},\ and\ \citenamefont
  {Mahler}}]{Allahverdyan2010}%
  \BibitemOpen
  \bibfield  {author} {\bibinfo {author} {\bibfnamefont {A.~E.}\ \bibnamefont
  {Allahverdyan}}, \bibinfo {author} {\bibfnamefont {K.}~\bibnamefont
  {Hovhannisyan}}, \ and\ \bibinfo {author} {\bibfnamefont {G.}~\bibnamefont
  {Mahler}},\ }\href@noop {} {\bibfield  {journal} {\bibinfo  {journal} {Phys.
  Rev. E}\ }\textbf {\bibinfo {volume} {81}},\ \bibinfo {pages} {051129}
  (\bibinfo {year} {2010})}\BibitemShut {NoStop}%
\bibitem [{\citenamefont {Apertet}\ \emph {et~al.}(2013)\citenamefont
  {Apertet}, \citenamefont {Ouerdane}, \citenamefont {Michot}, \citenamefont
  {Goupil},\ and\ \citenamefont {Lecoeur}}]{Apertet2013A}%
  \BibitemOpen
  \bibfield  {author} {\bibinfo {author} {\bibfnamefont {Y.}~\bibnamefont
  {Apertet}}, \bibinfo {author} {\bibfnamefont {H.}~\bibnamefont {Ouerdane}},
  \bibinfo {author} {\bibfnamefont {A.}~\bibnamefont {Michot}}, \bibinfo
  {author} {\bibfnamefont {C.}~\bibnamefont {Goupil}}, \ and\ \bibinfo {author}
  {\bibfnamefont {P.}~\bibnamefont {Lecoeur}},\ }\href@noop {} {\bibfield
  {journal} {\bibinfo  {journal} {Europhys. Lett.}\ }\textbf {\bibinfo {volume}
  {103}},\ \bibinfo {pages} {40001} (\bibinfo {year} {2013})}\BibitemShut
  {NoStop}%
\bibitem [{\citenamefont {Curzon}\ and\ \citenamefont
  {Ahlborn}(1975)}]{CA1975}%
  \BibitemOpen
  \bibfield  {author} {\bibinfo {author} {\bibfnamefont {F.~L.}\ \bibnamefont
  {Curzon}}\ and\ \bibinfo {author} {\bibfnamefont {B.}~\bibnamefont
  {Ahlborn}},\ }\href@noop {} {\bibfield  {journal} {\bibinfo  {journal} {Am.
  J. Phys.}\ }\textbf {\bibinfo {volume} {43}},\ \bibinfo {pages} {22}
  (\bibinfo {year} {1975})}\BibitemShut {NoStop}%
\bibitem [{\citenamefont {Esposito}\ \emph {et~al.}(2010)\citenamefont
  {Esposito}, \citenamefont {Kawai}, \citenamefont {Lindenberg},\ and\
  \citenamefont {Van~den Broeck}}]{Esposito2010}%
  \BibitemOpen
  \bibfield  {author} {\bibinfo {author} {\bibfnamefont {M.}~\bibnamefont
  {Esposito}}, \bibinfo {author} {\bibfnamefont {R.}~\bibnamefont {Kawai}},
  \bibinfo {author} {\bibfnamefont {K.}~\bibnamefont {Lindenberg}}, \ and\
  \bibinfo {author} {\bibfnamefont {C.}~\bibnamefont {Van~den Broeck}},\
  }\href@noop {} {\bibfield  {journal} {\bibinfo  {journal} {Phys. Rev. Lett.}\
  }\textbf {\bibinfo {volume} {105}},\ \bibinfo {pages} {150603} (\bibinfo
  {year} {2010})}\BibitemShut {NoStop}%
\bibitem [{\citenamefont {{Izumida, Y.}}\ and\ \citenamefont {{Okuda,
  K.}}(2012)}]{IzumidaOkuda}%
  \BibitemOpen
  \bibfield  {author} {\bibinfo {author} {\bibnamefont {{Izumida, Y.}}}\ and\
  \bibinfo {author} {\bibnamefont {{Okuda, K.}}},\ }\href@noop {} {\bibfield
  {journal} {\bibinfo  {journal} {Europhys. Lett.}\ }\textbf {\bibinfo {volume}
  {97}},\ \bibinfo {pages} {10004} (\bibinfo {year} {2012})}\BibitemShut
  {NoStop}%
\bibitem [{\citenamefont {Johal}(2010)}]{Sir2010}%
  \BibitemOpen
  \bibfield  {author} {\bibinfo {author} {\bibfnamefont {R.~S.}\ \bibnamefont
  {Johal}},\ }\href@noop {} {\bibfield  {journal} {\bibinfo  {journal} {Phys.
  Rev. E}\ }\textbf {\bibinfo {volume} {82}},\ \bibinfo {pages} {061113}
  (\bibinfo {year} {2010})}\BibitemShut {NoStop}%
\bibitem [{\citenamefont {Thomas}\ and\ \citenamefont {Johal}(2015)}]{GJ2015}%
  \BibitemOpen
  \bibfield  {author} {\bibinfo {author} {\bibfnamefont {G.}~\bibnamefont
  {Thomas}}\ and\ \bibinfo {author} {\bibfnamefont {R.~S.}\ \bibnamefont
  {Johal}},\ }\href@noop {} {\bibfield  {journal} {\bibinfo  {journal} {J.
  Phys. A}\ }\textbf {\bibinfo {volume} {48}},\ \bibinfo {pages} {335002}
  (\bibinfo {year} {2015})}\BibitemShut {NoStop}%
\bibitem [{\citenamefont {Johal}(2018)}]{Johal2018}%
  \BibitemOpen
  \bibfield  {author} {\bibinfo {author} {\bibfnamefont {R.~S.}\ \bibnamefont
  {Johal}},\ }\href@noop {} {\bibfield  {journal} {\bibinfo  {journal}
  {Europhys. Lett.}\ }\textbf {\bibinfo {volume} {121}},\ \bibinfo {pages}
  {50009} (\bibinfo {year} {2018})}\BibitemShut {NoStop}%
\bibitem [{\citenamefont {de~Tom\'as}\ \emph {et~al.}(2012)\citenamefont
  {de~Tom\'as}, \citenamefont {Hern\'andez},\ and\ \citenamefont
  {Roco}}]{deTomas2012}%
  \BibitemOpen
  \bibfield  {author} {\bibinfo {author} {\bibfnamefont {C.}~\bibnamefont
  {de~Tom\'as}}, \bibinfo {author} {\bibfnamefont {A.~C.}\ \bibnamefont
  {Hern\'andez}}, \ and\ \bibinfo {author} {\bibfnamefont {J.~M.~M.}\
  \bibnamefont {Roco}},\ }\href@noop {} {\bibfield  {journal} {\bibinfo
  {journal} {Phys. Rev. E}\ }\textbf {\bibinfo {volume} {85}},\ \bibinfo
  {pages} {010104} (\bibinfo {year} {2012})}\BibitemShut {NoStop}%
\bibitem [{\citenamefont {Izumida}\ \emph {et~al.}(2013)\citenamefont
  {Izumida}, \citenamefont {Okuda}, \citenamefont {Hern{\'a}ndez},\ and\
  \citenamefont {Roco}}]{Izumida2013}%
  \BibitemOpen
  \bibfield  {author} {\bibinfo {author} {\bibfnamefont {Y.}~\bibnamefont
  {Izumida}}, \bibinfo {author} {\bibfnamefont {K.}~\bibnamefont {Okuda}},
  \bibinfo {author} {\bibfnamefont {A.~C.}\ \bibnamefont {Hern{\'a}ndez}}, \
  and\ \bibinfo {author} {\bibfnamefont {J.}~\bibnamefont {Roco}},\ }\href@noop
  {} {\bibfield  {journal} {\bibinfo  {journal} {Europhys. Lett.}\ }\textbf
  {\bibinfo {volume} {101}},\ \bibinfo {pages} {10005} (\bibinfo {year}
  {2013})}\BibitemShut {NoStop}%
\bibitem [{\citenamefont {Velasco}\ \emph {et~al.}(1997)\citenamefont
  {Velasco}, \citenamefont {Roco}, \citenamefont {Medina},\ and\ \citenamefont
  {Hern\'andez}}]{Velasco1997}%
  \BibitemOpen
  \bibfield  {author} {\bibinfo {author} {\bibfnamefont {S.}~\bibnamefont
  {Velasco}}, \bibinfo {author} {\bibfnamefont {J.~M.~M.}\ \bibnamefont
  {Roco}}, \bibinfo {author} {\bibfnamefont {A.}~\bibnamefont {Medina}}, \ and\
  \bibinfo {author} {\bibfnamefont {A.~C.}\ \bibnamefont {Hern\'andez}},\
  }\href@noop {} {\bibfield  {journal} {\bibinfo  {journal} {Phys. Rev. Lett.}\
  }\textbf {\bibinfo {volume} {78}},\ \bibinfo {pages} {3241} (\bibinfo {year}
  {1997})}\BibitemShut {NoStop}%
\bibitem [{\citenamefont {Johal}(2019)}]{Johal2019}%
  \BibitemOpen
  \bibfield  {author} {\bibinfo {author} {\bibfnamefont {R.~S.}\ \bibnamefont
  {Johal}},\ }\href@noop {} {\bibfield  {journal} {\bibinfo  {journal}
  {arXiv:1906.02453}\ } (\bibinfo {year} {2019})}\BibitemShut {NoStop}%
\bibitem [{\citenamefont {Abah}\ and\ \citenamefont {Lutz}(2016)}]{LutzEPL}%
  \BibitemOpen
  \bibfield  {author} {\bibinfo {author} {\bibfnamefont {O.}~\bibnamefont
  {Abah}}\ and\ \bibinfo {author} {\bibfnamefont {E.}~\bibnamefont {Lutz}},\
  }\href@noop {} {\bibfield  {journal} {\bibinfo  {journal} {Europhys. Lett.}\
  }\textbf {\bibinfo {volume} {113}},\ \bibinfo {pages} {60002} (\bibinfo
  {year} {2016})}\BibitemShut {NoStop}%
\bibitem [{\citenamefont {Correa}\ \emph
  {et~al.}(2014{\natexlab{a}})\citenamefont {Correa}, \citenamefont {Palao},
  \citenamefont {Adesso},\ and\ \citenamefont {Alonso}}]{Alonso2014B}%
  \BibitemOpen
  \bibfield  {author} {\bibinfo {author} {\bibfnamefont {L.~A.}\ \bibnamefont
  {Correa}}, \bibinfo {author} {\bibfnamefont {J.~P.}\ \bibnamefont {Palao}},
  \bibinfo {author} {\bibfnamefont {G.}~\bibnamefont {Adesso}}, \ and\ \bibinfo
  {author} {\bibfnamefont {D.}~\bibnamefont {Alonso}},\ }\href@noop {}
  {\bibfield  {journal} {\bibinfo  {journal} {Phys. Rev. E}\ }\textbf {\bibinfo
  {volume} {90}},\ \bibinfo {pages} {062124} (\bibinfo {year}
  {2014}{\natexlab{a}})}\BibitemShut {NoStop}%
\bibitem [{\citenamefont {Scovil}\ and\ \citenamefont
  {Schulz-DuBois}(1959)}]{Scovil1959}%
  \BibitemOpen
  \bibfield  {author} {\bibinfo {author} {\bibfnamefont {H.~E.~D.}\
  \bibnamefont {Scovil}}\ and\ \bibinfo {author} {\bibfnamefont {E.~O.}\
  \bibnamefont {Schulz-DuBois}},\ }\href@noop {} {\bibfield  {journal}
  {\bibinfo  {journal} {Phys. Rev. Lett.}\ }\textbf {\bibinfo {volume} {2}},\
  \bibinfo {pages} {262} (\bibinfo {year} {1959})}\BibitemShut {NoStop}%
\bibitem [{\citenamefont {Geusic}\ \emph {et~al.}(1959)\citenamefont {Geusic},
  \citenamefont {Schulz-DuBois}, \citenamefont {De~Grasse},\ and\ \citenamefont
  {Scovil}}]{Scovil1959B}%
  \BibitemOpen
  \bibfield  {author} {\bibinfo {author} {\bibfnamefont {J.~E.}\ \bibnamefont
  {Geusic}}, \bibinfo {author} {\bibfnamefont {E.~O.}\ \bibnamefont
  {Schulz-DuBois}}, \bibinfo {author} {\bibfnamefont {R.~W.}\ \bibnamefont
  {De~Grasse}}, \ and\ \bibinfo {author} {\bibfnamefont {H.~E.~D.}\
  \bibnamefont {Scovil}},\ }\href@noop {} {\bibfield  {journal} {\bibinfo
  {journal} {J. Appl. Phys.}\ }\textbf {\bibinfo {volume} {30}},\ \bibinfo
  {pages} {1113} (\bibinfo {year} {1959})}\BibitemShut {NoStop}%
\bibitem [{\citenamefont {Kosloff}(2013)}]{KosloffEntropy}%
  \BibitemOpen
  \bibfield  {author} {\bibinfo {author} {\bibfnamefont {R.}~\bibnamefont
  {Kosloff}},\ }\href@noop {} {\bibfield  {journal} {\bibinfo  {journal}
  {Entropy}\ }\textbf {\bibinfo {volume} {15}},\ \bibinfo {pages} {2100}
  (\bibinfo {year} {2013})}\BibitemShut {NoStop}%
\bibitem [{\citenamefont {Mahler}(2014)}]{Mahler}%
  \BibitemOpen
  \bibfield  {author} {\bibinfo {author} {\bibfnamefont {G.}~\bibnamefont
  {Mahler}},\ }\href@noop {} {\emph {\bibinfo {title} {Quantum thermodynamic
  processes: Energy and information flow at the nanoscale}}}\ (\bibinfo
  {publisher} {Jenny Stanford Publishing},\ \bibinfo {year} {2014})\BibitemShut
  {NoStop}%
\bibitem [{\citenamefont {Vinjanampathy}\ and\ \citenamefont
  {Anders}(2016)}]{SV2016}%
  \BibitemOpen
  \bibfield  {author} {\bibinfo {author} {\bibfnamefont {S.}~\bibnamefont
  {Vinjanampathy}}\ and\ \bibinfo {author} {\bibfnamefont {J.}~\bibnamefont
  {Anders}},\ }\href {\doibase 10.1080/00107514.2016.1201896} {\bibfield
  {journal} {\bibinfo  {journal} {Contemp. Phys.}\ }\textbf {\bibinfo {volume}
  {57}},\ \bibinfo {pages} {545} (\bibinfo {year} {2016})}\BibitemShut
  {NoStop}%
\bibitem [{\citenamefont {Millen}\ and\ \citenamefont {Xuereb}(2016)}]{Xuereb}%
  \BibitemOpen
  \bibfield  {author} {\bibinfo {author} {\bibfnamefont {J.}~\bibnamefont
  {Millen}}\ and\ \bibinfo {author} {\bibfnamefont {A.}~\bibnamefont
  {Xuereb}},\ }\href@noop {} {\bibfield  {journal} {\bibinfo  {journal} {New J.
  Phys.}\ }\textbf {\bibinfo {volume} {18}},\ \bibinfo {pages} {011002}
  (\bibinfo {year} {2016})}\BibitemShut {NoStop}%
\bibitem [{\citenamefont {Deffner}\ and\ \citenamefont
  {Campbell}(2019)}]{DeffnerBook}%
  \BibitemOpen
  \bibfield  {author} {\bibinfo {author} {\bibfnamefont {S.}~\bibnamefont
  {Deffner}}\ and\ \bibinfo {author} {\bibfnamefont {S.}~\bibnamefont
  {Campbell}},\ }\href@noop {} {\emph {\bibinfo {title} {Quantum
  Thermodynamics}}}\ (\bibinfo  {publisher} {Morgan \& Claypool Publishers},\
  \bibinfo {year} {2019})\BibitemShut {NoStop}%
\bibitem [{\citenamefont {Alicki}\ and\ \citenamefont
  {Kosloff}(2018)}]{AlickiKosloff}%
  \BibitemOpen
  \bibfield  {author} {\bibinfo {author} {\bibfnamefont {R.}~\bibnamefont
  {Alicki}}\ and\ \bibinfo {author} {\bibfnamefont {R.}~\bibnamefont
  {Kosloff}},\ }in\ \href@noop {} {\emph {\bibinfo {booktitle} {Thermodynamics
  in the Quantum Regime}}}\ (\bibinfo  {publisher} {Springer},\ \bibinfo {year}
  {2018})\ pp.\ \bibinfo {pages} {1--33}\BibitemShut {NoStop}%
\bibitem [{\citenamefont {Binder}\ \emph {et~al.}(2018)\citenamefont {Binder},
  \citenamefont {Correa}, \citenamefont {Gogolin}, \citenamefont {Anders},\
  and\ \citenamefont {Adesso}}]{Binder}%
  \BibitemOpen
  \bibfield  {author} {\bibinfo {author} {\bibfnamefont {F.}~\bibnamefont
  {Binder}}, \bibinfo {author} {\bibfnamefont {L.~A.}\ \bibnamefont {Correa}},
  \bibinfo {author} {\bibfnamefont {C.}~\bibnamefont {Gogolin}}, \bibinfo
  {author} {\bibfnamefont {J.}~\bibnamefont {Anders}}, \ and\ \bibinfo {author}
  {\bibfnamefont {G.}~\bibnamefont {Adesso}},\ }\href@noop {} {\emph {\bibinfo
  {title} {Thermodynamics in the Quantum Regime: Fundamental Aspects and New
  Directions}}}\ (\bibinfo  {publisher} {Springer},\ \bibinfo {year}
  {2018})\BibitemShut {NoStop}%
\bibitem [{\citenamefont {Geva}\ and\ \citenamefont
  {Kosloff}(1994)}]{Geva1994}%
  \BibitemOpen
  \bibfield  {author} {\bibinfo {author} {\bibfnamefont {E.}~\bibnamefont
  {Geva}}\ and\ \bibinfo {author} {\bibfnamefont {R.}~\bibnamefont {Kosloff}},\
  }\href@noop {} {\bibfield  {journal} {\bibinfo  {journal} {Phys. Rev. E}\
  }\textbf {\bibinfo {volume} {49}},\ \bibinfo {pages} {3903} (\bibinfo {year}
  {1994})}\BibitemShut {NoStop}%
\bibitem [{\citenamefont {Geva}\ and\ \citenamefont
  {Kosloff}(1996)}]{Geva1996}%
  \BibitemOpen
  \bibfield  {author} {\bibinfo {author} {\bibfnamefont {E.}~\bibnamefont
  {Geva}}\ and\ \bibinfo {author} {\bibfnamefont {R.}~\bibnamefont {Kosloff}},\
  }\href@noop {} {\bibfield  {journal} {\bibinfo  {journal} {J. Chem. Phys.}\
  }\textbf {\bibinfo {volume} {104}},\ \bibinfo {pages} {7681} (\bibinfo {year}
  {1996})}\BibitemShut {NoStop}%
\bibitem [{\citenamefont {Kosloff}\ and\ \citenamefont
  {Levy}(2014)}]{LevyKosloff}%
  \BibitemOpen
  \bibfield  {author} {\bibinfo {author} {\bibfnamefont {R.}~\bibnamefont
  {Kosloff}}\ and\ \bibinfo {author} {\bibfnamefont {A.}~\bibnamefont {Levy}},\
  }\href@noop {} {\bibfield  {journal} {\bibinfo  {journal} {Annu. Rev. Phys.
  Chem.}\ }\textbf {\bibinfo {volume} {65}},\ \bibinfo {pages} {365} (\bibinfo
  {year} {2014})}\BibitemShut {NoStop}%
\bibitem [{\citenamefont {Dorfman}\ \emph {et~al.}(2018)\citenamefont
  {Dorfman}, \citenamefont {Xu},\ and\ \citenamefont {Cao}}]{Dorfman2018}%
  \BibitemOpen
  \bibfield  {author} {\bibinfo {author} {\bibfnamefont {K.~E.}\ \bibnamefont
  {Dorfman}}, \bibinfo {author} {\bibfnamefont {D.}~\bibnamefont {Xu}}, \ and\
  \bibinfo {author} {\bibfnamefont {J.}~\bibnamefont {Cao}},\ }\href@noop {}
  {\bibfield  {journal} {\bibinfo  {journal} {Phys. Rev. E}\ }\textbf {\bibinfo
  {volume} {97}},\ \bibinfo {pages} {042120} (\bibinfo {year}
  {2018})}\BibitemShut {NoStop}%
\bibitem [{\citenamefont {Singh}\ and\ \citenamefont {Johal}(2019)}]{VJ2019}%
  \BibitemOpen
  \bibfield  {author} {\bibinfo {author} {\bibfnamefont {V.}~\bibnamefont
  {Singh}}\ and\ \bibinfo {author} {\bibfnamefont {R.~S.}\ \bibnamefont
  {Johal}},\ }\href@noop {} {\bibfield  {journal} {\bibinfo  {journal} {Phys.
  Rev. E}\ }\textbf {\bibinfo {volume} {100}},\ \bibinfo {pages} {012138}
  (\bibinfo {year} {2019})}\BibitemShut {NoStop}%
\bibitem [{\citenamefont {Jaseem}\ \emph {et~al.}(2018)\citenamefont {Jaseem},
  \citenamefont {Hajdu{\v{s}}ek}, \citenamefont {Vedral}, \citenamefont
  {Fazio}, \citenamefont {Kwek},\ and\ \citenamefont
  {Vinjanampathy}}]{Jaseem2019}%
  \BibitemOpen
  \bibfield  {author} {\bibinfo {author} {\bibfnamefont {N.}~\bibnamefont
  {Jaseem}}, \bibinfo {author} {\bibfnamefont {M.}~\bibnamefont
  {Hajdu{\v{s}}ek}}, \bibinfo {author} {\bibfnamefont {V.}~\bibnamefont
  {Vedral}}, \bibinfo {author} {\bibfnamefont {R.}~\bibnamefont {Fazio}},
  \bibinfo {author} {\bibfnamefont {L.~C.}\ \bibnamefont {Kwek}}, \ and\
  \bibinfo {author} {\bibfnamefont {S.}~\bibnamefont {Vinjanampathy}},\
  }\href@noop {} {\bibfield  {journal} {\bibinfo  {journal} {arXiv:1812.10082}\
  } (\bibinfo {year} {2018})}\BibitemShut {NoStop}%
\bibitem [{\citenamefont {Klatzow}\ \emph {et~al.}(2019)\citenamefont
  {Klatzow}, \citenamefont {Becker}, \citenamefont {Ledingham}, \citenamefont
  {Weinzetl}, \citenamefont {Kaczmarek}, \citenamefont {Saunders},
  \citenamefont {Nunn}, \citenamefont {Walmsley}, \citenamefont {Uzdin},\ and\
  \citenamefont {Poem}}]{Uzdin2019}%
  \BibitemOpen
  \bibfield  {author} {\bibinfo {author} {\bibfnamefont {J.}~\bibnamefont
  {Klatzow}}, \bibinfo {author} {\bibfnamefont {J.~N.}\ \bibnamefont {Becker}},
  \bibinfo {author} {\bibfnamefont {P.~M.}\ \bibnamefont {Ledingham}}, \bibinfo
  {author} {\bibfnamefont {C.}~\bibnamefont {Weinzetl}}, \bibinfo {author}
  {\bibfnamefont {K.~T.}\ \bibnamefont {Kaczmarek}}, \bibinfo {author}
  {\bibfnamefont {D.~J.}\ \bibnamefont {Saunders}}, \bibinfo {author}
  {\bibfnamefont {J.}~\bibnamefont {Nunn}}, \bibinfo {author} {\bibfnamefont
  {I.~A.}\ \bibnamefont {Walmsley}}, \bibinfo {author} {\bibfnamefont
  {R.}~\bibnamefont {Uzdin}}, \ and\ \bibinfo {author} {\bibfnamefont
  {E.}~\bibnamefont {Poem}},\ }\href@noop {} {\bibfield  {journal} {\bibinfo
  {journal} {Phys. Rev. Lett.}\ }\textbf {\bibinfo {volume} {122}},\ \bibinfo
  {pages} {110601} (\bibinfo {year} {2019})}\BibitemShut {NoStop}%
\bibitem [{\citenamefont {Linden}\ \emph {et~al.}(2010)\citenamefont {Linden},
  \citenamefont {Popescu},\ and\ \citenamefont {Skrzypczyk}}]{Linden2010}%
  \BibitemOpen
  \bibfield  {author} {\bibinfo {author} {\bibfnamefont {N.}~\bibnamefont
  {Linden}}, \bibinfo {author} {\bibfnamefont {S.}~\bibnamefont {Popescu}}, \
  and\ \bibinfo {author} {\bibfnamefont {P.}~\bibnamefont {Skrzypczyk}},\
  }\href@noop {} {\bibfield  {journal} {\bibinfo  {journal} {Phys. Rev. Lett.}\
  }\textbf {\bibinfo {volume} {105}},\ \bibinfo {pages} {130401} (\bibinfo
  {year} {2010})}\BibitemShut {NoStop}%
\bibitem [{\citenamefont {Levy}\ and\ \citenamefont
  {Kosloff}(2012)}]{Levy2012}%
  \BibitemOpen
  \bibfield  {author} {\bibinfo {author} {\bibfnamefont {A.}~\bibnamefont
  {Levy}}\ and\ \bibinfo {author} {\bibfnamefont {R.}~\bibnamefont {Kosloff}},\
  }\href@noop {} {\bibfield  {journal} {\bibinfo  {journal} {Phys. Rev. Lett.}\
  }\textbf {\bibinfo {volume} {108}},\ \bibinfo {pages} {070604} (\bibinfo
  {year} {2012})}\BibitemShut {NoStop}%
\bibitem [{\citenamefont {Correa}\ \emph {et~al.}(2013)\citenamefont {Correa},
  \citenamefont {Palao}, \citenamefont {Adesso},\ and\ \citenamefont
  {Alonso}}]{Alonso2013}%
  \BibitemOpen
  \bibfield  {author} {\bibinfo {author} {\bibfnamefont {L.~A.}\ \bibnamefont
  {Correa}}, \bibinfo {author} {\bibfnamefont {J.~P.}\ \bibnamefont {Palao}},
  \bibinfo {author} {\bibfnamefont {G.}~\bibnamefont {Adesso}}, \ and\ \bibinfo
  {author} {\bibfnamefont {D.}~\bibnamefont {Alonso}},\ }\href@noop {}
  {\bibfield  {journal} {\bibinfo  {journal} {Phys. Rev. E}\ }\textbf {\bibinfo
  {volume} {87}},\ \bibinfo {pages} {042131} (\bibinfo {year}
  {2013})}\BibitemShut {NoStop}%
\bibitem [{\citenamefont {Agarwalla}\ \emph {et~al.}(2017)\citenamefont
  {Agarwalla}, \citenamefont {Jiang},\ and\ \citenamefont {Segal}}]{Bijay2017}%
  \BibitemOpen
  \bibfield  {author} {\bibinfo {author} {\bibfnamefont {B.~K.}\ \bibnamefont
  {Agarwalla}}, \bibinfo {author} {\bibfnamefont {J.-H.}\ \bibnamefont
  {Jiang}}, \ and\ \bibinfo {author} {\bibfnamefont {D.}~\bibnamefont
  {Segal}},\ }\href@noop {} {\bibfield  {journal} {\bibinfo  {journal} {Phys.
  Rev. B}\ }\textbf {\bibinfo {volume} {96}},\ \bibinfo {pages} {104304}
  (\bibinfo {year} {2017})}\BibitemShut {NoStop}%
\bibitem [{\citenamefont {Kilgour}\ and\ \citenamefont
  {Segal}(2018)}]{Segal2018}%
  \BibitemOpen
  \bibfield  {author} {\bibinfo {author} {\bibfnamefont {M.}~\bibnamefont
  {Kilgour}}\ and\ \bibinfo {author} {\bibfnamefont {D.}~\bibnamefont
  {Segal}},\ }\href@noop {} {\bibfield  {journal} {\bibinfo  {journal} {Phys.
  Rev. E}\ }\textbf {\bibinfo {volume} {98}},\ \bibinfo {pages} {012117}
  (\bibinfo {year} {2018})}\BibitemShut {NoStop}%
\bibitem [{\citenamefont {Holubec}\ and\ \citenamefont
  {Novotný}(2019)}]{Holubec2019}%
  \BibitemOpen
  \bibfield  {author} {\bibinfo {author} {\bibfnamefont {V.}~\bibnamefont
  {Holubec}}\ and\ \bibinfo {author} {\bibfnamefont {T.}~\bibnamefont
  {Novotný}},\ }\href@noop {} {\bibfield  {journal} {\bibinfo  {journal} {J.
  Chem. Phys.}\ }\textbf {\bibinfo {volume} {151}},\ \bibinfo {pages} {044108}
  (\bibinfo {year} {2019})}\BibitemShut {NoStop}%
\bibitem [{\citenamefont {Maslennikov}\ \emph {et~al.}(2019)\citenamefont
  {Maslennikov}, \citenamefont {Ding}, \citenamefont {Habl{\"u}tzel},
  \citenamefont {Gan}, \citenamefont {Roulet}, \citenamefont {Nimmrichter},
  \citenamefont {Dai}, \citenamefont {Scarani},\ and\ \citenamefont
  {Matsukevich}}]{Scarani2019}%
  \BibitemOpen
  \bibfield  {author} {\bibinfo {author} {\bibfnamefont {G.}~\bibnamefont
  {Maslennikov}}, \bibinfo {author} {\bibfnamefont {S.}~\bibnamefont {Ding}},
  \bibinfo {author} {\bibfnamefont {R.}~\bibnamefont {Habl{\"u}tzel}}, \bibinfo
  {author} {\bibfnamefont {J.}~\bibnamefont {Gan}}, \bibinfo {author}
  {\bibfnamefont {A.}~\bibnamefont {Roulet}}, \bibinfo {author} {\bibfnamefont
  {S.}~\bibnamefont {Nimmrichter}}, \bibinfo {author} {\bibfnamefont
  {J.}~\bibnamefont {Dai}}, \bibinfo {author} {\bibfnamefont {V.}~\bibnamefont
  {Scarani}}, \ and\ \bibinfo {author} {\bibfnamefont {D.}~\bibnamefont
  {Matsukevich}},\ }\href@noop {} {\bibfield  {journal} {\bibinfo  {journal}
  {Nat. Commun.}\ }\textbf {\bibinfo {volume} {10}},\ \bibinfo {pages} {202}
  (\bibinfo {year} {2019})}\BibitemShut {NoStop}%
\bibitem [{\citenamefont {Mitchison}\ \emph {et~al.}(2016)\citenamefont
  {Mitchison}, \citenamefont {Huber}, \citenamefont {Prior}, \citenamefont
  {Woods},\ and\ \citenamefont {Plenio}}]{Mitchison2016}%
  \BibitemOpen
  \bibfield  {author} {\bibinfo {author} {\bibfnamefont {M.~T.}\ \bibnamefont
  {Mitchison}}, \bibinfo {author} {\bibfnamefont {M.}~\bibnamefont {Huber}},
  \bibinfo {author} {\bibfnamefont {J.}~\bibnamefont {Prior}}, \bibinfo
  {author} {\bibfnamefont {M.~P.}\ \bibnamefont {Woods}}, \ and\ \bibinfo
  {author} {\bibfnamefont {M.~B.}\ \bibnamefont {Plenio}},\ }\href@noop {}
  {\bibfield  {journal} {\bibinfo  {journal} {Quantum Science and Technology}\
  }\textbf {\bibinfo {volume} {1}},\ \bibinfo {pages} {015001} (\bibinfo {year}
  {2016})}\BibitemShut {NoStop}%
\bibitem [{\citenamefont {Brask}\ and\ \citenamefont
  {Brunner}(2015)}]{Brunner2015}%
  \BibitemOpen
  \bibfield  {author} {\bibinfo {author} {\bibfnamefont {J.~B.}\ \bibnamefont
  {Brask}}\ and\ \bibinfo {author} {\bibfnamefont {N.}~\bibnamefont
  {Brunner}},\ }\href@noop {} {\bibfield  {journal} {\bibinfo  {journal} {Phys.
  Rev. E}\ }\textbf {\bibinfo {volume} {92}},\ \bibinfo {pages} {062101}
  (\bibinfo {year} {2015})}\BibitemShut {NoStop}%
\bibitem [{\citenamefont {Gorini}\ \emph {et~al.}(1976)\citenamefont {Gorini},
  \citenamefont {Kossakowski},\ and\ \citenamefont {Sudarshan}}]{Gorini}%
  \BibitemOpen
  \bibfield  {author} {\bibinfo {author} {\bibfnamefont {V.}~\bibnamefont
  {Gorini}}, \bibinfo {author} {\bibfnamefont {A.}~\bibnamefont {Kossakowski}},
  \ and\ \bibinfo {author} {\bibfnamefont {E.~C.~G.}\ \bibnamefont
  {Sudarshan}},\ }\href@noop {} {\bibfield  {journal} {\bibinfo  {journal} {J.
  Math. Phys.}\ }\textbf {\bibinfo {volume} {17}},\ \bibinfo {pages} {821}
  (\bibinfo {year} {1976})}\BibitemShut {NoStop}%
\bibitem [{\citenamefont {Lindblad}(1976)}]{Lindblad}%
  \BibitemOpen
  \bibfield  {author} {\bibinfo {author} {\bibfnamefont {G.}~\bibnamefont
  {Lindblad}},\ }\href@noop {} {\bibfield  {journal} {\bibinfo  {journal}
  {Commun. Math. Phys.}\ }\textbf {\bibinfo {volume} {48}},\ \bibinfo {pages}
  {119} (\bibinfo {year} {1976})}\BibitemShut {NoStop}%
\bibitem [{\citenamefont {Boukobza}\ and\ \citenamefont
  {Tannor}(2006{\natexlab{a}})}]{BoukobzaTannor2006A}%
  \BibitemOpen
  \bibfield  {author} {\bibinfo {author} {\bibfnamefont {E.}~\bibnamefont
  {Boukobza}}\ and\ \bibinfo {author} {\bibfnamefont {D.~J.}\ \bibnamefont
  {Tannor}},\ }\href@noop {} {\bibfield  {journal} {\bibinfo  {journal} {Phys.
  Rev. A}\ }\textbf {\bibinfo {volume} {74}},\ \bibinfo {pages} {063823}
  (\bibinfo {year} {2006}{\natexlab{a}})}\BibitemShut {NoStop}%
\bibitem [{\citenamefont {Boukobza}\ and\ \citenamefont
  {Tannor}(2006{\natexlab{b}})}]{BoukobzaTannor2006B}%
  \BibitemOpen
  \bibfield  {author} {\bibinfo {author} {\bibfnamefont {E.}~\bibnamefont
  {Boukobza}}\ and\ \bibinfo {author} {\bibfnamefont {D.~J.}\ \bibnamefont
  {Tannor}},\ }\href@noop {} {\bibfield  {journal} {\bibinfo  {journal} {Phys.
  Rev. A}\ }\textbf {\bibinfo {volume} {74}},\ \bibinfo {pages} {063822}
  (\bibinfo {year} {2006}{\natexlab{b}})}\BibitemShut {NoStop}%
\bibitem [{\citenamefont {Boukobza}\ and\ \citenamefont
  {Tannor}(2007)}]{BoukobzaTannor2007}%
  \BibitemOpen
  \bibfield  {author} {\bibinfo {author} {\bibfnamefont {E.}~\bibnamefont
  {Boukobza}}\ and\ \bibinfo {author} {\bibfnamefont {D.~J.}\ \bibnamefont
  {Tannor}},\ }\href@noop {} {\bibfield  {journal} {\bibinfo  {journal} {Phys.
  Rev. Lett.}\ }\textbf {\bibinfo {volume} {98}},\ \bibinfo {pages} {240601}
  (\bibinfo {year} {2007})}\BibitemShut {NoStop}%
\bibitem [{\citenamefont {Alicki}(1979)}]{Alicki1979}%
  \BibitemOpen
  \bibfield  {author} {\bibinfo {author} {\bibfnamefont {R.}~\bibnamefont
  {Alicki}},\ }\href@noop {} {\bibfield  {journal} {\bibinfo  {journal} {J.
  Phys. A}\ }\textbf {\bibinfo {volume} {12}},\ \bibinfo {pages} {L103}
  (\bibinfo {year} {1979})}\BibitemShut {NoStop}%
\bibitem [{\citenamefont {Kosloff}(1984)}]{Kosloff1984}%
  \BibitemOpen
  \bibfield  {author} {\bibinfo {author} {\bibfnamefont {R.}~\bibnamefont
  {Kosloff}},\ }\href@noop {} {\bibfield  {journal} {\bibinfo  {journal} {J.
  Chem. Phys.}\ }\textbf {\bibinfo {volume} {80}},\ \bibinfo {pages} {1625}
  (\bibinfo {year} {1984})}\BibitemShut {NoStop}%
\bibitem [{\citenamefont {Geva}\ and\ \citenamefont
  {Kosloff}(1992)}]{GevaKosloff1992}%
  \BibitemOpen
  \bibfield  {author} {\bibinfo {author} {\bibfnamefont {E.}~\bibnamefont
  {Geva}}\ and\ \bibinfo {author} {\bibfnamefont {R.}~\bibnamefont {Kosloff}},\
  }\href@noop {} {\bibfield  {journal} {\bibinfo  {journal} {J. Chem. Phys.}\
  }\textbf {\bibinfo {volume} {96}},\ \bibinfo {pages} {3054} (\bibinfo {year}
  {1992})}\BibitemShut {NoStop}%
\bibitem [{\citenamefont {Abah}\ \emph {et~al.}(2012)\citenamefont {Abah},
  \citenamefont {Ro\ss{}nagel}, \citenamefont {Jacob}, \citenamefont {Deffner},
  \citenamefont {Schmidt-Kaler}, \citenamefont {Singer},\ and\ \citenamefont
  {Lutz}}]{Lutz2012}%
  \BibitemOpen
  \bibfield  {author} {\bibinfo {author} {\bibfnamefont {O.}~\bibnamefont
  {Abah}}, \bibinfo {author} {\bibfnamefont {J.}~\bibnamefont {Ro\ss{}nagel}},
  \bibinfo {author} {\bibfnamefont {G.}~\bibnamefont {Jacob}}, \bibinfo
  {author} {\bibfnamefont {S.}~\bibnamefont {Deffner}}, \bibinfo {author}
  {\bibfnamefont {F.}~\bibnamefont {Schmidt-Kaler}}, \bibinfo {author}
  {\bibfnamefont {K.}~\bibnamefont {Singer}}, \ and\ \bibinfo {author}
  {\bibfnamefont {E.}~\bibnamefont {Lutz}},\ }\href@noop {} {\bibfield
  {journal} {\bibinfo  {journal} {Phys. Rev. Lett.}\ }\textbf {\bibinfo
  {volume} {109}},\ \bibinfo {pages} {203006} (\bibinfo {year}
  {2012})}\BibitemShut {NoStop}%
\bibitem [{\citenamefont {Correa}\ \emph
  {et~al.}(2014{\natexlab{b}})\citenamefont {Correa}, \citenamefont {Palao},
  \citenamefont {Alonso},\ and\ \citenamefont {Adesso}}]{Alonso2014}%
  \BibitemOpen
  \bibfield  {author} {\bibinfo {author} {\bibfnamefont {L.~A.}\ \bibnamefont
  {Correa}}, \bibinfo {author} {\bibfnamefont {J.~P.}\ \bibnamefont {Palao}},
  \bibinfo {author} {\bibfnamefont {D.}~\bibnamefont {Alonso}}, \ and\ \bibinfo
  {author} {\bibfnamefont {G.}~\bibnamefont {Adesso}},\ }\href@noop {}
  {\bibfield  {journal} {\bibinfo  {journal} {Sci. Rep}\ }\textbf {\bibinfo
  {volume} {4}},\ \bibinfo {pages} {3949} (\bibinfo {year}
  {2014}{\natexlab{b}})}\BibitemShut {NoStop}%
\bibitem [{\citenamefont {Uzdin}\ and\ \citenamefont
  {Kosloff}(2014)}]{UzdinEPL}%
  \BibitemOpen
  \bibfield  {author} {\bibinfo {author} {\bibfnamefont {R.}~\bibnamefont
  {Uzdin}}\ and\ \bibinfo {author} {\bibfnamefont {R.}~\bibnamefont
  {Kosloff}},\ }\href@noop {} {\bibfield  {journal} {\bibinfo  {journal}
  {Europhys. Lett.}\ }\textbf {\bibinfo {volume} {108}},\ \bibinfo {pages}
  {40001} (\bibinfo {year} {2014})}\BibitemShut {NoStop}%
\bibitem [{\citenamefont {Izumida}\ \emph {et~al.}(2015)\citenamefont
  {Izumida}, \citenamefont {Okuda}, \citenamefont {Roco},\ and\ \citenamefont
  {Hern\'andez}}]{Izumida2015}%
  \BibitemOpen
  \bibfield  {author} {\bibinfo {author} {\bibfnamefont {Y.}~\bibnamefont
  {Izumida}}, \bibinfo {author} {\bibfnamefont {K.}~\bibnamefont {Okuda}},
  \bibinfo {author} {\bibfnamefont {J.~M.~M.}\ \bibnamefont {Roco}}, \ and\
  \bibinfo {author} {\bibfnamefont {A.~C.}\ \bibnamefont {Hern\'andez}},\
  }\href {\doibase 10.1103/PhysRevE.91.052140} {\bibfield  {journal} {\bibinfo
  {journal} {Phys. Rev. E}\ }\textbf {\bibinfo {volume} {91}},\ \bibinfo
  {pages} {052140} (\bibinfo {year} {2015})}\BibitemShut {NoStop}%
\bibitem [{\citenamefont {Yuan}\ \emph {et~al.}(2014)\citenamefont {Yuan},
  \citenamefont {Wang}, \citenamefont {He}, \citenamefont {Ma},\ and\
  \citenamefont {Wang}}]{Yuan2014}%
  \BibitemOpen
  \bibfield  {author} {\bibinfo {author} {\bibfnamefont {Y.}~\bibnamefont
  {Yuan}}, \bibinfo {author} {\bibfnamefont {R.}~\bibnamefont {Wang}}, \bibinfo
  {author} {\bibfnamefont {J.}~\bibnamefont {He}}, \bibinfo {author}
  {\bibfnamefont {Y.}~\bibnamefont {Ma}}, \ and\ \bibinfo {author}
  {\bibfnamefont {J.}~\bibnamefont {Wang}},\ }\href@noop {} {\bibfield
  {journal} {\bibinfo  {journal} {Phys. Rev. E}\ }\textbf {\bibinfo {volume}
  {90}},\ \bibinfo {pages} {052151} (\bibinfo {year} {2014})}\BibitemShut
  {NoStop}%
\bibitem [{\citenamefont {Wang}\ \emph {et~al.}(2012)\citenamefont {Wang},
  \citenamefont {Li}, \citenamefont {Tu}, \citenamefont {Hern\'andez},\ and\
  \citenamefont {Roco}}]{WangLi2012}%
  \BibitemOpen
  \bibfield  {author} {\bibinfo {author} {\bibfnamefont {Y.}~\bibnamefont
  {Wang}}, \bibinfo {author} {\bibfnamefont {M.}~\bibnamefont {Li}}, \bibinfo
  {author} {\bibfnamefont {Z.~C.}\ \bibnamefont {Tu}}, \bibinfo {author}
  {\bibfnamefont {A.~C.}\ \bibnamefont {Hern\'andez}}, \ and\ \bibinfo {author}
  {\bibfnamefont {J.~M.~M.}\ \bibnamefont {Roco}},\ }\href@noop {} {\bibfield
  {journal} {\bibinfo  {journal} {Phys. Rev. E}\ }\textbf {\bibinfo {volume}
  {86}},\ \bibinfo {pages} {011127} (\bibinfo {year} {2012})}\BibitemShut
  {NoStop}%
\bibitem [{\citenamefont {Kleiner}(2007)}]{Kleiner2007}%
  \BibitemOpen
  \bibfield  {author} {\bibinfo {author} {\bibfnamefont {I.}~\bibnamefont
  {Kleiner}},\ }in\ \href@noop {} {\emph {\bibinfo {booktitle} {A History of
  Abstract Algebra}}}\ (\bibinfo  {publisher} {Springer},\ \bibinfo {year}
  {2007})\ pp.\ \bibinfo {pages} {113--163}\BibitemShut {NoStop}%
\bibitem [{\citenamefont {Stewart}(1990)}]{Stewart1990}%
  \BibitemOpen
  \bibfield  {author} {\bibinfo {author} {\bibfnamefont {I.}~\bibnamefont
  {Stewart}},\ }\href@noop {} {\emph {\bibinfo {title} {Galois theory}}}\
  (\bibinfo  {publisher} {Chapman and Hall/CRC},\ \bibinfo {year}
  {1990})\BibitemShut {NoStop}%
\bibitem [{\citenamefont {Feynman}\ \emph {et~al.}(2008)\citenamefont
  {Feynman}, \citenamefont {Leighton},\ and\ \citenamefont {Sands}}]{Feynman}%
  \BibitemOpen
  \bibfield  {author} {\bibinfo {author} {\bibfnamefont {R.~P.}\ \bibnamefont
  {Feynman}}, \bibinfo {author} {\bibfnamefont {R.~B.}\ \bibnamefont
  {Leighton}}, \ and\ \bibinfo {author} {\bibfnamefont {M.}~\bibnamefont
  {Sands}},\ }\href@noop {} {\emph {\bibinfo {title} {The Feynman Lectures on
  Physics}}}\ (\bibinfo  {publisher} {Narosa Publishing House},\ \bibinfo
  {address} {New Delhi, India},\ \bibinfo {year} {2008})\BibitemShut {NoStop}%
\bibitem [{\citenamefont {Parrondo}\ and\ \citenamefont
  {Espanol}(1996)}]{Parrondo}%
  \BibitemOpen
  \bibfield  {author} {\bibinfo {author} {\bibfnamefont {J.~M.~R.}\
  \bibnamefont {Parrondo}}\ and\ \bibinfo {author} {\bibfnamefont
  {P.}~\bibnamefont {Espanol}},\ }\href@noop {} {\bibfield  {journal} {\bibinfo
   {journal} {Am. J. Phys.}\ }\textbf {\bibinfo {volume} {64}},\ \bibinfo
  {pages} {1125} (\bibinfo {year} {1996})}\BibitemShut {NoStop}%
\bibitem [{\citenamefont {Sheng}\ \emph {et~al.}(2014)\citenamefont {Sheng},
  \citenamefont {Yang},\ and\ \citenamefont {Tu}}]{ShengYangFR}%
  \BibitemOpen
  \bibfield  {author} {\bibinfo {author} {\bibfnamefont {S.}~\bibnamefont
  {Sheng}}, \bibinfo {author} {\bibfnamefont {P.}~\bibnamefont {Yang}}, \ and\
  \bibinfo {author} {\bibfnamefont {Z.~C.}\ \bibnamefont {Tu}},\ }\href@noop {}
  {\bibfield  {journal} {\bibinfo  {journal} {Commun. Theor. Phys.}\ }\textbf
  {\bibinfo {volume} {62}},\ \bibinfo {pages} {589} (\bibinfo {year}
  {2014})}\BibitemShut {NoStop}%
\bibitem [{\citenamefont {Singh}\ and\ \citenamefont
  {Johal}(2017)}]{VarinderJohal}%
  \BibitemOpen
  \bibfield  {author} {\bibinfo {author} {\bibfnamefont {V.}~\bibnamefont
  {Singh}}\ and\ \bibinfo {author} {\bibfnamefont {R.~S.}\ \bibnamefont
  {Johal}},\ }\href@noop {} {\bibfield  {journal} {\bibinfo  {journal}
  {Entropy}\ }\textbf {\bibinfo {volume} {19}},\ \bibinfo {pages} {576}
  (\bibinfo {year} {2017})}\BibitemShut {NoStop}%
\bibitem [{\citenamefont {Singh}\ and\ \citenamefont
  {Johal}(2018)}]{VarinderJohal2018}%
  \BibitemOpen
  \bibfield  {author} {\bibinfo {author} {\bibfnamefont {V.}~\bibnamefont
  {Singh}}\ and\ \bibinfo {author} {\bibfnamefont {R.~S.}\ \bibnamefont
  {Johal}},\ }\href@noop {} {\bibfield  {journal} {\bibinfo  {journal} {J.
  Stat. Mech.}\ }\textbf {\bibinfo {volume} {2018}},\ \bibinfo {pages} {073205}
  (\bibinfo {year} {2018})}\BibitemShut {NoStop}%
\bibitem [{\citenamefont {Mani}\ and\ \citenamefont
  {Benjamin}(2019)}]{Benjamin}%
  \BibitemOpen
  \bibfield  {author} {\bibinfo {author} {\bibfnamefont {A.}~\bibnamefont
  {Mani}}\ and\ \bibinfo {author} {\bibfnamefont {C.}~\bibnamefont
  {Benjamin}},\ }\href@noop {} {\bibfield  {journal} {\bibinfo  {journal} {J.
  Phys. Chem. C}\ }\textbf {\bibinfo {volume} {123}},\ \bibinfo {pages} {22858}
  (\bibinfo {year} {2019})}\BibitemShut {NoStop}%
  \end{thebibliography}
%\bibliographystyle{apsrev4-1}

\end{document}